\documentclass[10pt,a4paper,twoside]{report}
\usepackage{amsmath}
\usepackage[numbers,sort&compress]{natbib} 

\usepackage{notoccite}

\usepackage[utf8]{inputenc}   

\usepackage[english]{babel} 

\newcommand{\acknowledgments}{@undefined} 
\addto\captionsenglish{\renewcommand{\acknowledgments}{Acknowledgments}}

\addto\captionsfrench{\renewcommand{\acknowledgments}{Remerciements}}
\addto\captionsfrench{} 

\addto\captionsportuguese{\renewcommand{\acknowledgments}{Agradecimentos}}
\addto\captionsportuguese{} 

\newcommand{\coverThesis}{@undefined} 
\newcommand{\coverSupervisors}{@undefined} 
\newcommand{\coverExaminationCommittee}{@undefined} 
\newcommand{\coverChairperson}{@undefined} 
\newcommand{\coverSupervisor}{@undefined} 
\newcommand{\coverMemberCommittee}{@undefined} 
\addto\captionsenglish{\renewcommand{\coverThesis}{Thesis to obtain the Master of Science Degree in}}
\addto\captionsenglish{\renewcommand{\coverSupervisors}{Supervisor(s)}}
\addto\captionsenglish{\renewcommand{\coverExaminationCommittee}{Examination Committee}}
\addto\captionsenglish{\renewcommand{\coverChairperson}{Chairperson}}
\addto\captionsenglish{\renewcommand{\coverSupervisor}{Supervisor}}
\addto\captionsenglish{\renewcommand{\coverMemberCommittee}{Member of the Committee}}
\addto\captionsfrench{\renewcommand{\coverThesis}{Th\`ese pour l'obtention du Maîtrise des Sciences en}}
\addto\captionsfrench{\renewcommand{\coverSupervisors}{Directeur(s) de th\`ese}}
\addto\captionsfrench{\renewcommand{\coverExaminationCommittee}{Jury}}
\addto\captionsfrench{\renewcommand{\coverChairperson}{Pr\'esident}}
\addto\captionsfrench{\renewcommand{\coverSupervisor}{Directeur de th\`ese}}
\addto\captionsfrench{\renewcommand{\coverMemberCommittee}{Rapporteur}}
\addto\captionsportuguese{\renewcommand{\coverThesis}{Disserta\c{c}\~{a}o para obten\c{c}\~{a}o do Grau de Mestre em}}
\addto\captionsportuguese{\renewcommand{\coverSupervisors}{Orientador(es)}}
\addto\captionsportuguese{\renewcommand{\coverExaminationCommittee}{J\'{u}ri}}
\addto\captionsportuguese{\renewcommand{\coverChairperson}{Presidente}}
\addto\captionsportuguese{\renewcommand{\coverSupervisor}{Orientador}}
\addto\captionsportuguese{\renewcommand{\coverMemberCommittee}{Vogal}}

\newcommand{\declarationTitle}{@undefined} 
\newcommand{\declarationText}{@undefined}  
\addto\captionsenglish{\renewcommand{\declarationTitle}{Declaration}}
\addto\captionsenglish{\renewcommand{\declarationText}{I declare that this document is an original work of my own authorship and that it fulfills all the requirements of the Code of Conduct and Good Practices of the Universidade de Lisboa.}}
\addto\captionsportuguese{\renewcommand{\declarationTitle}{Declara\c{c}\~{a}o}}
\addto\captionsportuguese{\renewcommand{\declarationText}{Declaro que o presente documento \'{e} um trabalho original da minha autoria e que cumpre todos os requisitos do C\'{o}digo de Conduta e Boas Pr\'{a}ticas da Universidade de Lisboa.}}

\def\FontLb{
  \usefont{T1}{phv}{b}{n}\fontsize{16pt}{16pt}\selectfont}

\def\FontMb{
  \usefont{T1}{phv}{b}{n}\fontsize{14pt}{14pt}\selectfont}
\def\FontSn{
  \usefont{T1}{phv}{m}{n}\fontsize{12pt}{12pt}\selectfont}

\usepackage{geometry}	
\usepackage{setspace}

\usepackage{indentfirst}

\usepackage{graphicx}

\usepackage{amsmath}  
\usepackage{amsthm}   
\usepackage{amsfonts} %

\usepackage{subfigmat}

\usepackage{dcolumn}
\newcolumntype{d}{D{.}{.}{-1}} 
\newcolumntype{e}{D{E}{E}{-1}} 

\usepackage{rotating}

\usepackage{hyperref} 
\hypersetup{colorlinks,       
            linkcolor=black,  
            anchorcolor=black,
            citecolor=black,  
            filecolor=black,  
            menucolor=black,  
            urlcolor=black,   
	          bookmarksopen=false,    
	          bookmarksnumbered=true, 
	          pdftitle={Thesis},
            pdfauthor={Lízia Branco},
            pdfsubject={The role of ion channels in the transmissions of signals along axons},
            pdfkeywords={Na-K channels, Hodgkin-Huxley, Neuronal Spiking, Signal Propagation in the Axon, Action Potential Propagation Velocity, Diffusion Waves},
            pdfstartview=FitV,
            pdfdisplaydoctitle=true}

\usepackage[figure,table]{hypcap}

\usepackage{multirow}

\usepackage{booktabs}

\usepackage{pdfpages}

\usepackage{algorithm}
\usepackage{algpseudocode}

\begin{document}

\pagestyle{plain}

\pagenumbering{roman}

\thispagestyle {empty}

\includegraphics[bb=9.5cm 11cm 0cm 0cm,scale=0.29]{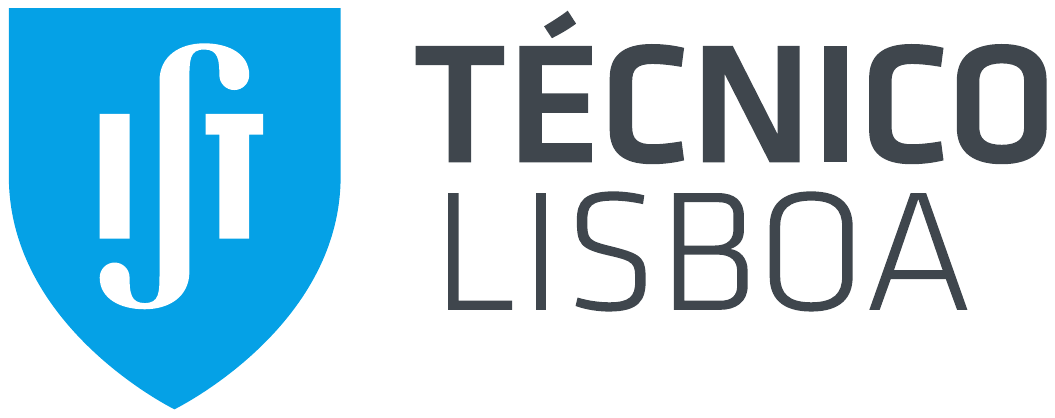}

\begin{center}
\vspace{2.5cm}
\vspace{1.0cm}
{\FontLb The role of ion channels in the\\ transmission of signals along axons} \\

\vspace{2.6cm}
{\FontMb Lízia Maria Gonçalves Branco} \\ 
\vspace{2.0cm}
{\FontSn \coverThesis} \\
\vspace{0.3cm}
{\FontLb Engineering Physics} \\ 
\vspace{1.0cm}
{\FontSn 
\begin{tabular}{ll}
 \coverSupervisors: Prof. Rui Manuel Agostinho Dil\~{a}o 
\end{tabular} } \\
\vspace{1.0cm}
{\FontMb \coverExaminationCommittee} \\
\vspace{0.3cm}
{\FontSn %
\begin{tabular}{c}
\coverChairperson:     Prof. Luís Humberto Viseu Melo  \\       
\coverSupervisor:      Prof. Rui Manuel Agostinho Dil\~{a}o \\ 
\coverMemberCommittee: Prof. Maria Teresa Ferreira Marques Pinheiro     
\end{tabular} } \\
\vspace{1.5cm}
{\FontMb July 2023} \\  examination

\end{center}

\cleardoublepage

\null\vskip5cm%
\begin{flushright}
     Dedicated to my family and closest friends, for inspiring and supporting me throughout this journey.
\end{flushright}
\vfill\newpage

\cleardoublepage

\null\vskip5cm
\begin{flushleft}
	\declarationTitle \\
	\declarationText
\end{flushleft}
\vfill\newpage

\cleardoublepage

\section*{\acknowledgments}

\addcontentsline{toc}{section}{\acknowledgments}

I would like to express my sincere gratitude and appreciation to the following individuals for their invaluable support and contributions throughout the completion of this thesis.

First and foremost, I am deeply indebted to my supervisor, Prof. Rui Dilão, for his guidance, expertise, and unwavering support. His insightful feedback, constructive criticism, and constant encouragement played a pivotal role in shaping the direction of this research.

I am indebted to my friends and family for their unconditional love, encouragement, and patience throughout this journey. Their unwavering belief in my abilities has been a constant source of motivation and inspiration.

Lastly, I would like to express my gratitude to all the individuals who provided support, advice, and encouragement during the course of this research. Your contributions have played an integral role in shaping the final outcome of this thesis.

I am truly honored to have had the opportunity to work on this thesis, and I am thankful to everyone who has contributed to its completion.

\cleardoublepage

\section*{Resumo}

\addcontentsline{toc}{section}{Resumo}
Neste trabalho, propomos um modelo eletrofisiológico bidimensional que descreve a resposta dos neurónios ao serem estimulados em condições de patch-clamp. Este modelo - uma versão reduzida do modelo de Hodgkin-Huxley - conserva as características essenciais da abordagem original, sendo, no entanto, mais eficiente e informativo.

Através da análise dos diagramas de bifurcação de diversos modelos reduzidos, descrevemos as contribuições dos diferentes canais iónicos na estabilidade ou instabilidade da membrana neuronal. Para além disso, exploramos a influência que a intensidade dos estímulos elétricos e vários parâmetros celulares têm na frequência de spiking dos neurónios.

Também nesta tese, apresentamos um modelo do axónio que simula a transmissão unidirecional de sinais e que não é afetado por perturbações no potencial da membrana. Após analisar o impacto da resistência intracelular e da capacitância membranar na velocidade do sinal, observamos discrepâncias consideráveis entre os nossos resultados e os previstos pela equação de cabo (\textit{cable equation model}).

A nossa investigação revela também que, quando a propagação do sinal ocorre sob a forma de múltiplos picos no potencial da membrana do axónio, há um transiente nos seus comprimentos de onda e frequências, sendo esse transiente influenciado pela resistência intracelular. Consequentemente, os picos têm velocidades distintas, sendo o primeiro pico mais rápido que os restantes. Concluímos também que a velocidade de propagação do primeiro pico - em alguns casos o único - se mantém relativamente constante ao longo do axónio, independentemente da intensidade do estímulo a que o neurónio está sujeito. O mesmo não se observa em picos subsequentes.

\vfill

\textbf{\Large Palavras-chave:} Canais Na-K, Modelo Hodgkin-Huxley, \textit{Spiking} Neuronal, Propagação de Sinais nos Axónios,   Ondas de Difusão.

\cleardoublepage

\section*{Abstract}

\addcontentsline{toc}{section}{Abstract}

This work proposes a two-dimensional electrophysiological model for describing neuronal responses to external electric stimuli under patch-clamped conditions. Our proposed model successfully captures the key features of the Hodgkin-Huxley model, while offering improved informativeness and numerical efficiency.

By analysing bifurcation diagrams from various reduced models, we describe the contributions of different ionic channels to the stability or instability of the neurones' response. We also explore the influence of various cell parameters in the neurone's spiking frequency, including the membrane capacitance and the ionic channels' conductivities and Nerst potentials.

Additionally, we suggest a spatial model for the axon that effectively simulates unidirectional signal transmission along neurons unaffected by membrane potential perturbations. After analysing the characteristics of this compartmental model, including the impact of intracellular resistance and membrane capacitance on the signal's velocity, we observed considerable discrepancies between our findings and those predicted by the cable equation model for signal propagation.

Moreover, our investigation reveals that when the transmission of neuronal signals occurs in multiple peaks, there is a transient variation in their wavelength and frequency, resulting in different peak velocities. Furthermore, our findings consistently demonstrate that the first peak - in some cases, the only peak observed - maintains a relatively constant velocity along the axon regardless of the stimulus intensity and is always faster than the subsequent peaks.

\vfill

\textbf{\Large Keywords:} Na-K channels, Hodgkin-Huxley Model, Neuronal Spiking, Signal Propagation in the Axon, Diffusion Waves.

\cleardoublepage

\setcounter{page}{1}
\pagenumbering{arabic}

\chapter{Thesis Outline}
\label{chapter:outline}

\section{Motivation}
\label{section:motivation}

Our body's regulation and proper functioning rely on the intricate exchange of ions between cells and the extracellular environment. Neurones, in particular, can transmit electrical signals throughout the body thanks to a controlled exchange of ions that triggers physical and chemical responses to internal and external stimuli. However, the mechanisms underlying this process are still controversial in the scientific community, which hardens our comprehension of equally relevant and often even more complex topics.

In our ever-evolving world, there are still many unanswered questions about the causes and possible treatments for some cerebral conditions, mostly because there is no fundamental knowledge about how our neurones work collectively in immensely complex networks. We are still incapable of understanding how our brain generates creativity or consciousness, for example, and the full potential of our neuronal system is yet to be deciphered. 

As we strive to unravel the mysteries of the brain, developing accurate and robust physical and mathematical models for signal propagation in neurones becomes increasingly important. These models offer a pathway to comprehend the complex dynamics of neuronal activity and thus promote revolutionary advancements in neuroscience and cognitive sciences. Moreover, the insights gained from studying neuronal networks provide the framework for developing advanced algorithms that may serve as the foundation for cutting-edge artificial intelligence systems.

This lack of insights into neuronal functioning motivated us to develop a novel physical model for signal transmission in neurones. While not the first in the quest to unravel the brain's secrets, our findings, if proven applicable compared to experimental data obtained in real neuronal systems, will bring us closer to a fundamentally strong understanding of neuronal activity and signal transmission. Although some questions remain unanswered, we might be closer to the truth now.

\section{Topic Overview}
\label{section:overview}

The neuronal membrane is a highly specialised structure. It consists of a thin lipid bilayer that is selectively permeable to certain ions, thus serving as a barrier between the intracellular and the extracellular space. Specialised proteins are embedded within the membrane to control the passage of ions in and out of the neurone. Depending on how they operate, these ion-specific channels can be categorised as passive or active. Passive channels allow ions to move across the membrane in response to a concentration gradient. On the other hand, active channels require energy input to operate and selectively allow ions to move across the membrane. The exchange of ions across the neuronal membrane is essential for many physiological functions, such as maintaining the resting membrane potential of the neurone, regulating the cell volume and transmitting electrical signals throughout the brain and the body.

In 1952, Alan Lloyd Hodgkin and Andrew Huxley \cite{Hodgkinquantitive}-\cite{Hodgkincomponents} presented a mathematical description of the neurones electrical activity and response to external electric stimuli. It soon became widely known as the Hodgkin-Huxley (HH) model and, owning the Nobel Prize in Physiology or Medicine in 1963, has led to a greater understanding of the mechanisms underlying action potential generation and signal propagation in neurones. This model consists of differential equations describing the time-dependent behaviour of three different ion-specific ion-specific channels: voltage-gated sodium (Na+) channels, voltage-gated potassium (K+) channels, and Leak channels. These equations, which later inspired the development of many other neuronal models, were based on the pioneer observation that the electrical activity of a squid giant axon was related to the opening and closing of ion channels, which inflicted changes in the membrane potential across the axon's membrane.

\section{Objectives and Deliverables}
\label{section:objectives}

Despite lacking theoretical support from physics and chemistry principles and being solely grounded by empirical observations in voltage clamp experiments, the HH model has shown good predictive power regarding the structural data of ionic channels. However, with its four coupled nonlinear equations and 25 parameters, the HH model hardens an intuitive understanding of the system's dynamics and is particularly inefficient when modelling neuronal networks. 

Intending to reduce the HH model's complexity and understand the precise role of each ionic channel in this empirical model, we conducted in-depth research on the model's features and proposed a new electrophysiological model for describing the neuronal response to external electric stimuli under patch-clamped conditions. 

Another topic that we focus on in this thesis relates to repetitive firing or spiking, which occurs when a neurone generates multiple action potentials in rapid succession in response to specific stimuli. Our objectives consisted of analysing the relation between the spiking frequency and the intensity of those stimuli and investigating how specific cell characteristics can modulate that frequency. Our findings facilitate further exploration of the dynamics of neuronal networks, synchronisation phenomena, and the generation of rhythmic behaviours in physiological and pathological conditions.

Although spatial uniformity can be achieved experimentally in vivo, under natural conditions the intricate structure of the neurone and the non-uniform distribution of charged bodies in the intracellular space create spatial gradients in the membrane potential. Therefore, we went beyond characterising local membrane responses and developed a spatial model for the axon that effectively simulates unidirectional signal transmission along neurones. We propose a new compartmental model based on the resistive properties of gap junctions and specific boundary conditions characterising the signals generated at the soma. Within this approach, we explore the influence that some parameters of the model have in the propagation of signals and describe in detail how resistive properties contribute to neuronal communication.

At this stage, it is important to note that our work offers concrete predictions regarding neuronal behaviour under several conditions that can be reproduced in vivo. This will allow our findings to be compared with experimental data obtained in real neuronal systems so that the applicability of our neuronal model can be assessed. 

\section{Thesis Outline}
\label{section:outline}

This work is divided into three chapters, with the present one being the first. Chapter 2 provides a concise theoretical overview of how the neuronal cellular membrane can be conceptualised as an electric circuit. This perspective allows us to comprehend the regulated movement of charged particles between the intracellular and extracellular space from a biological standpoint and through a physical and mathematical lens.

In Chapter 3, we analyse the membrane potential localised response to external stimuli. To better evaluate the relative importance of each channel and its specific role in the membrane's response to those stimuli, we tested different models where one, two, or a larger combination of ion-specific channels were present in the neuronal membrane. Subsequently, we focus on the dynamics of the variables utilised in the HH model. That allowed us to present a reduced physical model of the neuronal membrane, which employs only two independent variables: the membrane potential and one gating variable. We show that this model has the same dynamic properties as the original HH model with the advantage of having fewer parameters and a reduced phase space dimension.

Employing this reduced model, we then explore the impact of some of the membrane's parameters - such as ion channel conductance, Nernst potentials, and membrane capacitance - on the neuronal spiking behavior.

In the last section of Chapter 3, we focus on the global and spatially dependent membrane response to external stimuli and we present a spatial model for signal propagation along the axon. 

Finally,Chapter 4 encapsulates the findings derived from this study and outlines the potential future directions and prospects for this research.

\cleardoublepage

\chapter{The Hodgkin-Huxley Electrophysiological Model}
\label{chapter:background}

The potential difference between the extracellular and intracellular environments is called membrane potential, denoted by $V_m$. Due to a continuous exchange of ions between the interior and exterior environments, $V_m$ is time-dependent and can be described as $V_m(t) = V_i(t) - V_o(t)$, with $V_i$ and $V_o$ corresponding to the voltages inside and outside the cell, respectively.

The cell membrane, made up of a phospholipid bilayer, serves as a barrier between the interior and the exterior of the cells, functioning like a capacitor. By applying Coulomb's Law, we can derive the charge distribution across the membrane, denoted by $Q=C_mV_m$, where $C_m$ is the capacitance of the membrane. Similar to a real capacitor, a capacitive current, $I_C$, arises when the charge distribution across the membrane changes, and is given by $I_C = C_mdV_m/dt$. 

Due to various ion pumps and passive channels in the cell membrane, ions can move between the intracellular and extracellular spaces. This ion transport can occur through two main mechanisms: active transport, which requires energy in the form of ATP, and passive transport, which takes place without energy expenditure. Ion channels may be specific to a particular ion, and many of these channels can switch between an open and a closed state, allowing or blocking the passage of the respective ions. Their state depends on various factors, such as the membrane voltage and the concentration of various ions inside the neurone.

Due to the disparity in ion concentration between the inside and outside of the cell, the resulting concentration gradient across each ion-specific channel generates an electromotive force known as Nernst potential, $E_{ion}$. This force induces a flow of ions that depends on the conductivity $g_{ion}$ of that particular channel and that is given by $g_{ion}(V_m-E_{ion})$.

Because the membrane of neurones can present multiple ion-specific channels, the total ionic flow encompasses all individual contributions and the associated equation ruling such a system is
	\begin{equation}
	 I_C + I_{ion} = 0 \Leftrightarrow C\frac{dV_m}{dt} + \sum^{N_{ion}}_{j=1} g_{j,ion}(V_m-E_{j,ion}) = 0
  \label{eq-general}
	\end{equation}

\section{The Hodgkin and Huxley's Experiment}\label{section:2}

	In the late 1940s and early 1950s, Hodgkin and Huxley \cite{Hodgkinquantitive}-\cite{Hodgkincomponents} aimed at determining a physical model that could describe the response of the neuronal membrane to external electric stimuli. For that matter, they performed a series of experiments on the squid giant axon and, to guarantee that the membrane potential had the same value $V_m$ on the entire axon, they used a space clamp experimental technique to eliminate the spatial dependence of $V_m$. 

    Hodgkin and Huxley (HH) considered that the ionic channels acted independently and that the axon's membrane had three main channel types. Two ion-specific channels separated Sodium ($Na$) and Potassium ($K$). The remaining one was treated as a \textit{Leak} channel, through which other ions could pass. Therefore, the total ionic current in the HH model can be written as $I_{ion} = I_{Na}+I_{K}+I_{L}$, where $I_{Na}$ represents the ionic current in Sodium channels, $I_{K}$ to the ionic current in Potassium channels, and $I_{L}$ to the ionic current in Leak channels. Compared with the approach presented in the previous section and modelled by the equation \ref{eq-general}, Hodgkin and Huxley considered that the conductivities of the Potassium and Sodium channels were voltage-dependent. This implied that changes in $V_m$ would play a great role in the way $\text{Na}^{+}$ and $\text{K}^{+}$ moved between the intracellular and extracellular environment.
    
    To evaluate the axon's response to external stimuli, Hodgkin and Huxley subjected the axon's membrane to an externally controllable electric current $I$. Applying Kirchhoff's current law, the system can be described by $C_m dV_m/dt = I - (I_{Na}+I_{K}+I_{L})$.
    
	The recorded action potential voltage spikes were negative in the original Hodgkin and Huxley work published in 1952. However, in some contemporary conventions (\citet{rinzelmiller}), the membrane potential is considered to have the opposite sign. This convention results in action potential voltage spikes with positive values and Nerst potentials with the opposite sign compared to the original HH paper. It is important to note that these sign conventions are choices for ease of interpretation in different scientific contexts. The shift in sign convention does not affect the fundamental principles and insights provided by the Hodgkin-Huxley model but rather offers a different representation of the observed electrical phenomena. Throughout this thesis, our sign convention was the opposite of the HH original work, so the action potential voltage spikes are positive. 

	\subsection{Separation of Currents}
	Hodgkin and Huxley considered $C_m = 1$ $\mu F/$cm$^2$ and, under equilibrium conditions ($dV_m/dt = 0$, $I=0$), they determined the axon's resting potential, $V_r$. For convenience, they took the voltages relative to rest, i.e., $V = V_m-V_r$ and $V_i = E_i-V_r$ (with $i = \{Na, K, L\}$), and obtained $V_{Na} = 115$ mV and $V_{K} = -12$ mV. 
 
	Thanks to the voltage clamp experimental technique, Hodgkin and Huxley could hold $V_m$ fixed at (almost) any desired value. That allowed them to observe the response of $I_{ion}$ to the membrane potential change from a resting state to values that could mimic the action potential. As with $V_m$ fixed $I = I_{ion}$, $I$ was a direct measure of $I_{ion}$, being possible to trace $I_{ion}$ as a function of time.
    
	Describing the dependence of the three conductances of this model as a function of $V$ is not straightforward, as it requires the analysis of all the channels functioning autonomously and simultaneously, thus contributing, altogether, to the measured current $I$. With that in mind, Hodgkin and Huxley opted to determine $g_K(V)$ when $I_{Na} = 0$, a condition that could be accomplished by taking the Sodium ions out of the extracellular solution. Thus, from $I= I_{Na}+I_{K}+I_{L}$, they obtained a new current $I^{*}$ which was given by $I^{*} = I_{K}+I_{L}$. Thus, from $I$ and $I^{*}$, $I_{Na}$ could be studied independently using $I = I_{Na} - I^{*}$. 

    Because for $V = -84$ mV they noticed that changes in the external concentration of Potassium did not affect the current $I$, they concluded that $I_K(-84) = 0$ and determined a range of possible values for both $g_L$ and $V_L$: $g_L\in [0.13, 0.5]$ mS/$\text{cm}^2$ and $V_L\in [4, 22]$ mV. For other values of $V$, they could now use the previous results to determine $I_K(V)$ and, with that, $I_{Na}(V)$ using $I_K(V) = I^{*}(V) - g_L(V-V_L)$ and $I_{Na}(V) = I(V) - I^{*}(V)$.
	
	\subsection{Potassium Conductance}
	While analysing the behaviour of the Potassium conductance during a typical voltage clamp experiment - which can be written as $g_K(t) = I_K(t)/(V-V_K)$ - Hodgkin and Huxley proposed that
 	\[
	\begin{aligned}
    	&g_K = \bar{g}_K n^4(t) &&\quad && \frac{dn}{dt} =\alpha_n(V)(1-n)-\beta_n(V)n
    \end{aligned}
	\]
where $n \in [0,1]$ is a gating variable, $\bar{g}_K$ is a constant called maximal Potassium conductance, and $dn/dt$ follows the principle of mass action law. The power 4 in $n$ was chosen to fit the data best, although some researchers suggest that this particular value could be explained by the existence of four identical subunits on each Potassium channel that cooperatively control the passage of $K$ ions \cite{Jiang}, \cite{Kuang}.
	
	Naming the steady state of $n$ (obtained when $t \rightarrow \infty$) as $n_{\infty}$ and the exponential growth/decay constant in the voltage clamp experiments as $\tau_n(V)$, $dn/dt$ can also be written as
 	\[
	\begin{aligned}
    	\frac{dn}{dt} = \frac{n_{\infty}(V)-n}{\tau_n(V)} &&\quad && n_{\infty} = \frac{\alpha_n(V)}{\alpha_n(V)+\beta_n(V)} &&\quad && \tau_{n} = \frac{1}{\alpha_n(V)+\beta_n(V)}
    \end{aligned}
	\]

	Using data on $g_K$ from voltage clamp experiments with different values of $V$, Hodgkin and Huxley obtained $\bar{g}_K = 36$ mS/$\text{cm}^2$ and assumed the following fitting functions for $\alpha(V)$ and $\beta(V)$:
	 \[
	\begin{aligned}
    	\alpha_n(V) = \frac{0.01(10-V)}{e^{\frac{10-V}{10}}-1} &&\quad && \beta_n(V) = 0.125e^{-\frac{V}{80}}
    \end{aligned}
	\]
	
	\subsection{Sodium Conductance}
	By looking at the experimental data describing the time dependence of $g_{Na}$ for fixed membrane potential values, Hodgkin and Huxley proposed that 
	\[
	\begin{aligned}
    	g_{Na} = \bar{g}_{Na}m^3(t)h(t) &&\quad && \frac{\mathrm{d} m}{\mathrm{~d} t}=\alpha_m(V)(1-m)-\beta_m(V) m &&\quad && \frac{\mathrm{d} h}{\mathrm{~d} t}=\alpha_h(V)(1-h)-\beta_h(V) h
    \end{aligned}
	\]
where $m \in [0,1]$ is an activation variable, $h \in [0,1]$ an inactivation variable, $\bar{g}_{Na}$ corresponds to the maximal Sodium conductance, and $h$ and $m$ were assumed to obey the mass action law. Similarly to the activation variable of the Potassium, $dh/dt$ and $dm/dt$ can also be written as
	
	\[
	\begin{aligned}
    	\frac{\mathrm{d} m}{\mathrm{~d} t} = \frac{m_{\infty}(V)-m(V)}{\tau_m(V)}&&\quad && \frac{\mathrm{d} h}{\mathrm{~d} t} = \frac{h_{\infty}(V)-h(V)}{\tau_h(V)},
    \end{aligned}
	\]
where time constants $\tau_m$ and $\tau_h$ and the steady states $m_{\infty}$ and $h_{\infty}$ are defined in the same manner as $\tau_n$ and $n_{\infty}$, respectively.
	
    Once again assuming specific fitting functions for the mass action law parameters of $m$ and $h$, Hodgkin and Huxley obtained $\bar{g}_{Na}=120$ mS$/\text{cm}^2$ and 
    \[
	\begin{aligned}
    	\alpha_m(V)=\frac{0.1(-V+25)}{e^{\frac{-V+25}{10}}-1} &&\quad && \beta_m(V)=4 e^{\frac{-V}{18}}
    \end{aligned}
	\]

	\[
	\begin{aligned}
    	\alpha_h(V) &=0.07 e^{\frac{-V}{20}} &&\quad && \beta_h(V) &=\frac{1}{1+e^{\frac{-V+30}{10}}}
   	\end{aligned}
	\]

	\subsection{Leak Current}
	
	Considering $\bar{g}_L = 3$ mS/$\text{cm}^2$ and knowing that $0=I_{N a}(0)+I_K(0)+I_L(0) \land 0=g_{N a}\left(-V_{N a}\right)+g_K\left(-V_K\right)+g_L\left(-V_L\right)$ for $V=0$, Hodgkin and Huxley obtained $V_{L} = 10.613$ mV, using\[V_L=-(\bar{g}_{N a} m_{\infty}(0)^3 h_{\infty}(0) V_{N a}+\bar{g}_K h_{\infty}(0)^4)/g_L.\]
 
 \subsection{The Hodgkin-Huxley model}
\begin{figure}[!htb]
  	\centering
  	\includegraphics[width=0.5\textwidth]{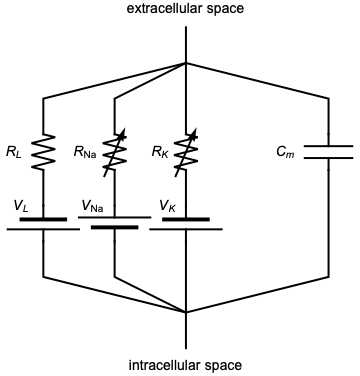}
  	\caption[Hodgkin Huxley model's electric circuit.]{Hodgkin Huxley model's electric circuit. We consider that the direction of positive current goes from the intracellular to the extracellular space.}
  	\label{fig:circuit}
	\end{figure}

	In summary, the HH model -- whose electric circuit is depicted in Figure \ref{fig:circuit} -- consists of the following four nonlinear ODEs:	

\begin{equation}
\begin{aligned}
\frac{dV}{dt} &=\frac{I}{C_m}-\frac{\bar{g}_{N a} m^3 h}{C_m}(V-V_{Na})\\
            &-\frac{\bar{g}_K }{C_m}n^4(V-V_K)-\frac{\bar{g}_L}{C_m}(V-V_L) \\
\frac{dm}{d t} &= \alpha_m(V)(1-m) - \beta_m(V)m \\
\frac{dh}{dt} &= \alpha_h(V)(1-h) - \beta_h(V)h \\
\frac{dn}{dt} &= \alpha_n(V)(1-n) - \beta_n(V)n
\end{aligned}
\end{equation}

Unless where specified, the values of the Hodgkin-Huxley model's parameters will remain unchanged throughout this work.

\section{Practical Considerations}

	It might be helpful to solve the previous equations and see what they can tell us about the response of neurones to localised electric stimulus $I$.

\begin{figure}[!htb]
  \begin{subfigmatrix}{2}
    \subfigure[Membrane potential response to two distinct electric stimuli applied from $t=50$ until $t=150$ ms.]{\includegraphics[width=0.6\linewidth]{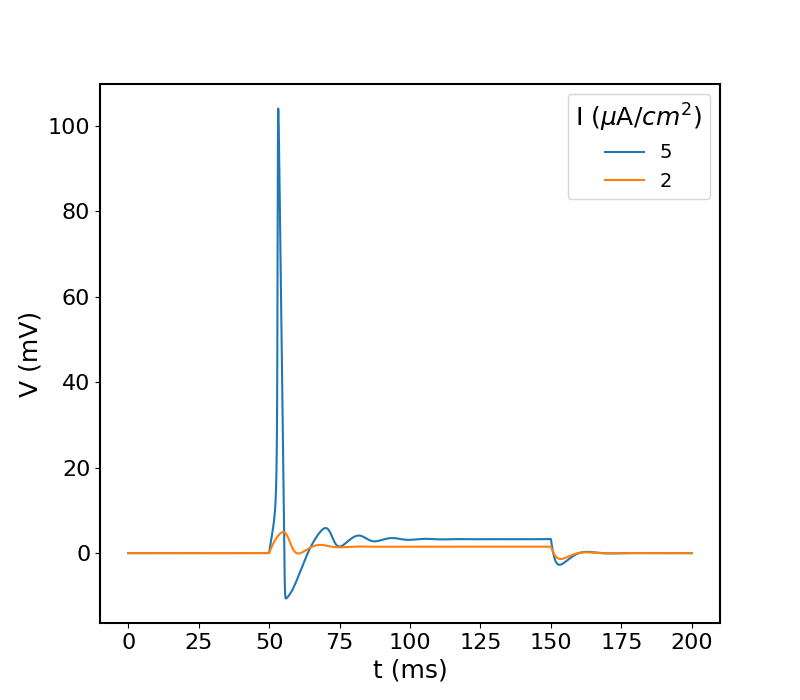}}
    \subfigure[Time evolution of the gating variables when an electric stimulus $I=5$ $\mu$A/$\text{cm}^2$ is turned on at $t=50$ and shut down at $t=150$ ms.]{\includegraphics[width=0.6\linewidth]{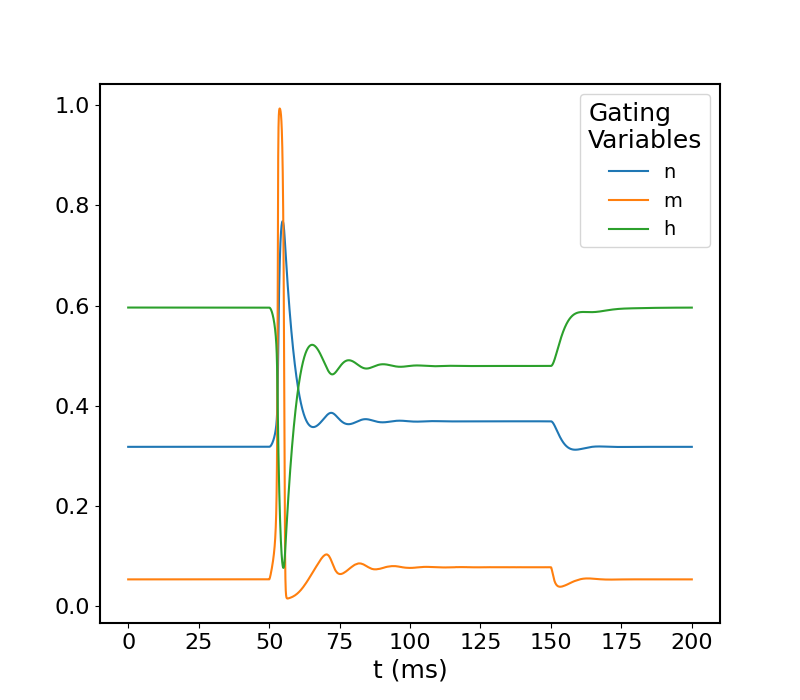}}
  \end{subfigmatrix}
  \caption[Action potentials' bell-shape curve and gating variables' response.]{Membranar response to distinct stimuli $I$. An action potentials' bell-shape curve is observed for $I=5$ $\mu$A/$\text{cm}^2$, which is accompanied by changes in the gating variables as observed in b).}
  \label{fig:bell-shape}
\end{figure}

	Considering that this soma signal is active from $t=50$ ms until $t = 150$ ms, the membrane potential evolves as represented in Figure \ref{fig:bell-shape} for $I= 2$ and $I = 5$ $\mu$A/$\text{cm}^2$. Here, we consider that the membrane is, initially, at its resting potential, so that $V_m=V_r \Leftrightarrow V = 0$.

	Hodgkin and Huxley presented models governed by the conductivity of multiple ion channels for studying a squid giant axon, but they soon proved useful in modelling many other excitable cells. One of the most important features of the HH model is the existence of a threshold electric stimulus for which an action potential -- the bell shape curve seen in Figure \ref{fig:bell-shape} a) for $I = 5$ $\mu$A/$\text{cm}^2$ -- is formed. The membrane potential quickly returns to equilibrium without significant fluctuations for input currents below this threshold. Instead, it rapidly converges to a stable state, effectively stopping the transmission of that stimulus, as seen in Figure \ref{fig:bell-shape} a) for $I = 2$ $\mu$A/$\text{cm}^2$. Each action potential begins with a rapid depolarisation of the membrane, followed by its repolarisation and hyperpolarisation. Each one of these events is accompanied and supported by changes in the gating variables $n$, $m$ and $h$, as represented in Figure \ref{fig:bell-shape} b).
 
\begin{figure}[!htb]
  \begin{subfigmatrix}{2}
    \subfigure[Zoomed-out view]{\includegraphics[width=0.6\linewidth]{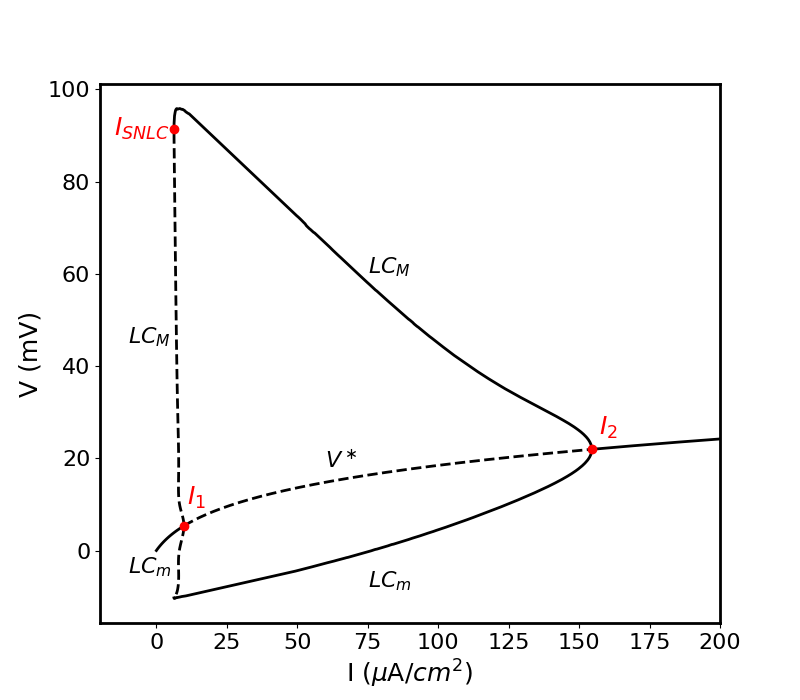}}
    \subfigure[Zoomed-in view]{\includegraphics[width=0.6\linewidth]{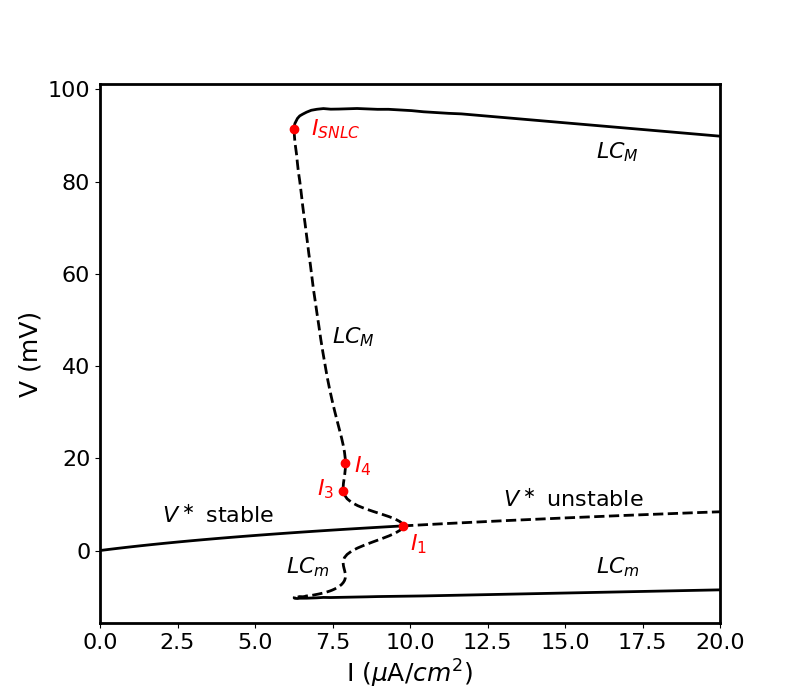}}
  \end{subfigmatrix}
  \caption[Bifurcation diagram of the Hodgkin-Huxley's electrophysiological model.]{Bifurcation diagram of the HH model as a function of the electric stimulus $I$, calculated with the software packages AUTO and XPPAUT.}
  \label{fig:auto-bif}
\end{figure}

    Using the bifurcation analysis software AUTO and XPPAUT \citet{Ermentrout}, we plotted in Figure \ref{fig:auto-bif} the steady state $V^{*}$ as a function of the electric stimulus $I$. Along the line $V^{*}$, there are two distinguishable states: a stable state, represented by the solid line, and an unstable state, represented by a dashed line. The two points where the transition between these two states occur correspond to Hopf bifurcations: the first being subcritical and occurring at $I_1 \approx 9.78$ $\mu$A/$\text{cm}^2$ and the second being supercritical and occurring at $I_2 \approx 154.52$ $\mu$A/$\text{cm}^2$. A limit cycle (LC) arises at these bifurcations, translating into spiking behaviour. Along the LC lines depicted in Figure \ref{fig:auto-bif}, the dashed line corresponds to the unstable periodic solutions, while the solid line corresponds to stable periodic cycles. The point where these two types of limit cycles coalesce is a saddle-node bifurcation of limit cycles (SNLC), which occurs at $I_{SNLC} \approx 6.26$ $\mu$A/$\text{cm}^2$. The maximum and minimum values of the membrane potential along the limit cycles are labelled as $LC_M$ and $LC_m$, respectively. The spiking phenomenon is observed in the range of stimuli $I_{SNLC} < I < I_2$. In Figure \ref{fig:auto-bif} b), the HH model's bifurcation diagram is depicted for a small range of $I$, enabling us to see that, for certain stimuli the HH equations can have two or three limit cycles. The chaotic behaviour investigated by \citet{Guckenheimer} occurs in the interval $[I_3 , I_4]$, right where HH equations have exactly three limit cycles.

\begin{figure}[!htb]
  \begin{subfigmatrix}{2}
    \subfigure[Time evolution of the membrane potentials for three different electric stimuli $I$ activated at $t=50$ ms and deactivated at $t=150$ ms.]{\includegraphics[width=0.6\linewidth]{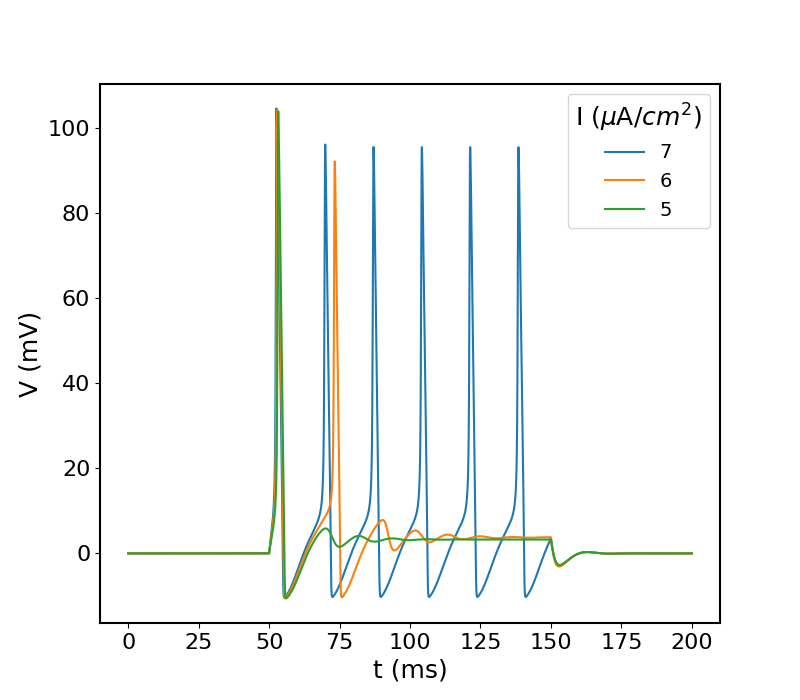}}
    \subfigure[Time evolution of the gating variables when the electric stimulus is set to $I=7$ $\mu$A/$\text{cm}^2$ from $t=50$ until $t=150$ ms.]{\includegraphics[width=0.6\linewidth]{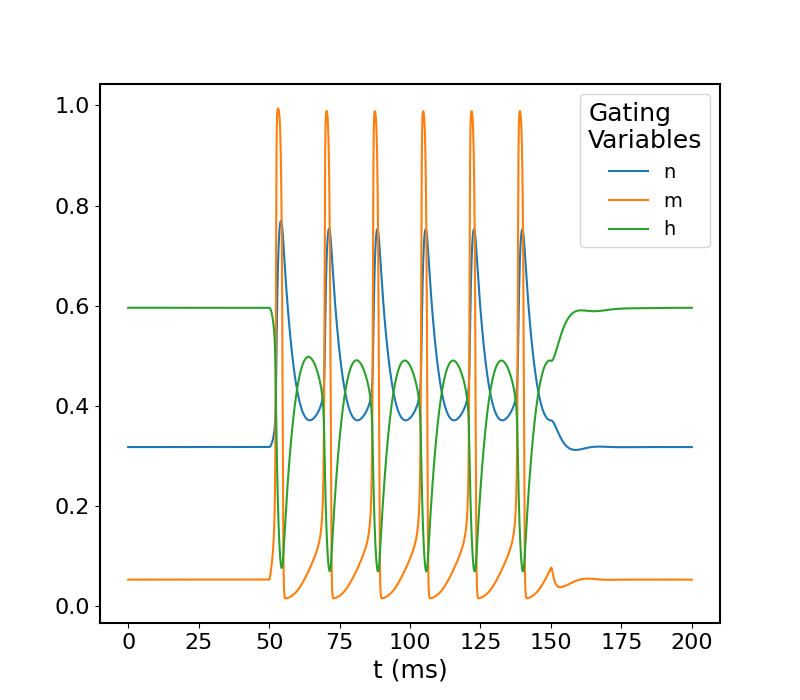}}
  \end{subfigmatrix}
  \caption[Distinct membrane responses to external stimuli.]{Distinct membrane responses to external stimuli. The cell membrane presents a unique action potential (green) for $I=5$ $\mu$A/$\text{cm}^2$, and intermittent response for $I=6$ $\mu$A/$\text{cm}^2$, and spiking activity for $I=7$ $\mu$A/$\text{cm}^2$.}
  \label{fig:time-evol}
\end{figure}

   $V^{*}$ is stable for $I<I_{1}$, and, in Figure \ref{fig:bell-shape}, we have observed two types of membrane responses to stimuli in that range. However, it is also possible for the membrane to exhibit an intermittent firing of action potentials. As shown by \citet{canogaspar}, this behaviour is due to a type I intermittency phenomenon near the codimension 2 SNLC bifurcation and the exact number of action potentials, $N$, is given by $\ln N=c-2\ln(I_{SNLC} - I)$, where $c$ is a constant. Also for $I<I_{1}$, the membrane can also present a spiking response, as long as $I_{SNLC} < I_1$. In the Figure \ref{fig:time-evol} a), we provide an example of intermittency for $I$ = 6 $\mu$A/$\text{cm}^2$ $<$ $I_{SNLC}$ $<$ $I_{1}$ and an example of spiking for $I$ = 7 $\mu$A/$\text{cm}^2$ $>$ $I_{SNLC}$ (which is also in the stable range). In Figure \ref{fig:time-evol} b), we represent the response of the gating variables when subjected to that exact stimulus $I$ = 7 $\mu$A/$\text{cm}^2$.

\cleardoublepage

\chapter{A new reduced Hodgkin-Huxley type model}
\label{chapter:implementation}

\section{Reduced Model Approach}

With four non-linear differential equations and several parameters, the HH equations are mathematically and computationally complex, posing challenges for an in-depth analysis and interpretation of neuronal dynamics.

Various approaches have been presented to reduce the HH model while maintaining its essential characteristics. One approach involves assuming certain gating variables to be constant or functions of the membrane potential. This simplification is often guided by the concept of slow and fast variables, where fast gating variables are assumed to adjust rapidly compared to other variables \cite{Rulkov2002ModelingOS}, \cite{fitzhugh1961445}.

We, on the other hand, building upon the work of \citet{canogaspar} -- who showed that the HH model without diffusion can be accurately described by the normal form of the Bautin bifurcation in polar coordinates --, were mainly concerned with testing several ways of reducing the complexity of the system while still capturing the essential features of the HH model's bifurcation diagram. Our rationale was that, as long as the essential features of the HH model's bifurcation diagram were preserved, the system's overall behaviour under external stimuli would remain intact.

\subsection{Contributions of the Ionic Channels in the HH Model}
\subsubsection{Bifurcation Analysis in Models with One and Two Types of Ionic Channels}

	For the first time, we want to fully understand the role played by each individual channel in the membrane's potential response to different stimuli. For that matter, we considered a set of models similar to the HH's but where the conductivity of two channels at a time was considered to be zero: a \textit{K model} (for a membrane with K channels only), a \textit{Na model} (for a membrane with Na channels only), and a \textit{Leak model} (for a membrane with Leak channels only). Then, we analysed the response of the cell membrane and the respective bifurcation diagram.
	
\begin{figure}[!htb]
  \begin{subfigmatrix}{2}
    \subfigure[Zoomed-in view]{\includegraphics[width=0.49\linewidth]{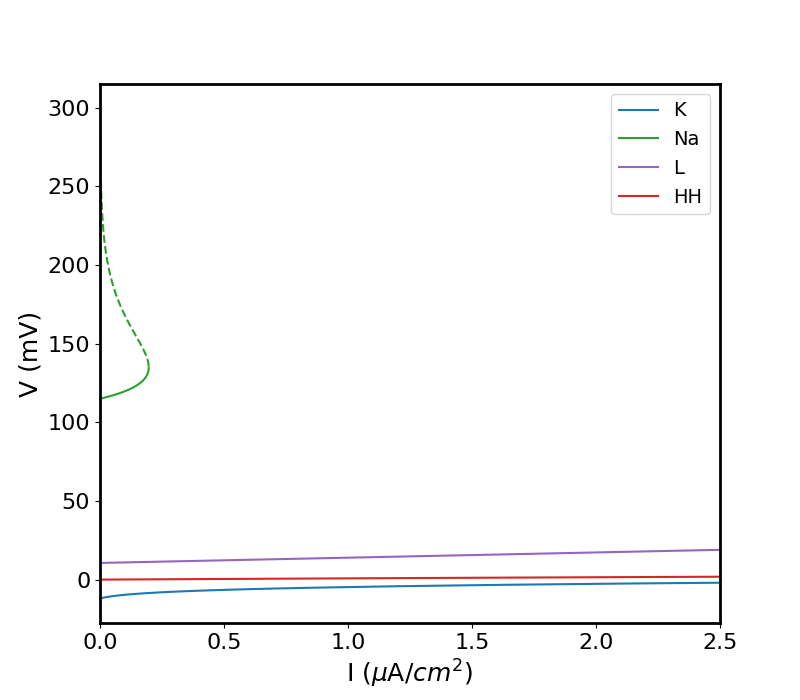}}
    \subfigure[Zoomed-out view]{\includegraphics[width=0.49\linewidth]{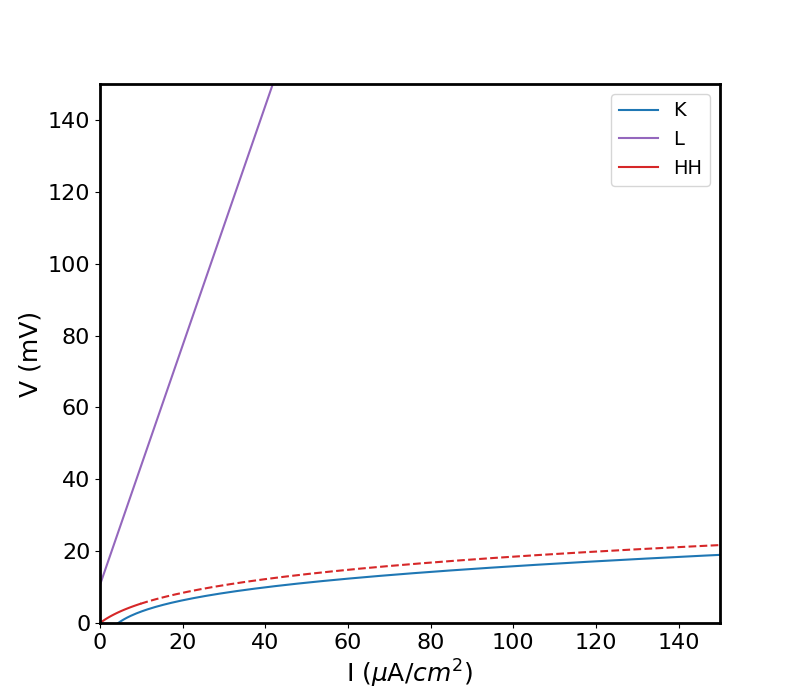}}
  \end{subfigmatrix}
  \caption[Location and stability of fixed points in electrophysiological models with a single ionic channel.]{Location and stability of the fixed points of electrophysiological models with a single ionic channel. As a reference, the fixed points of the Hodgkin-Huxley's model are depicted in red.}
  \label{fig:channels_na}
\end{figure}

\begin{figure}[!htb]
  \centering
  \includegraphics[width=\textwidth]{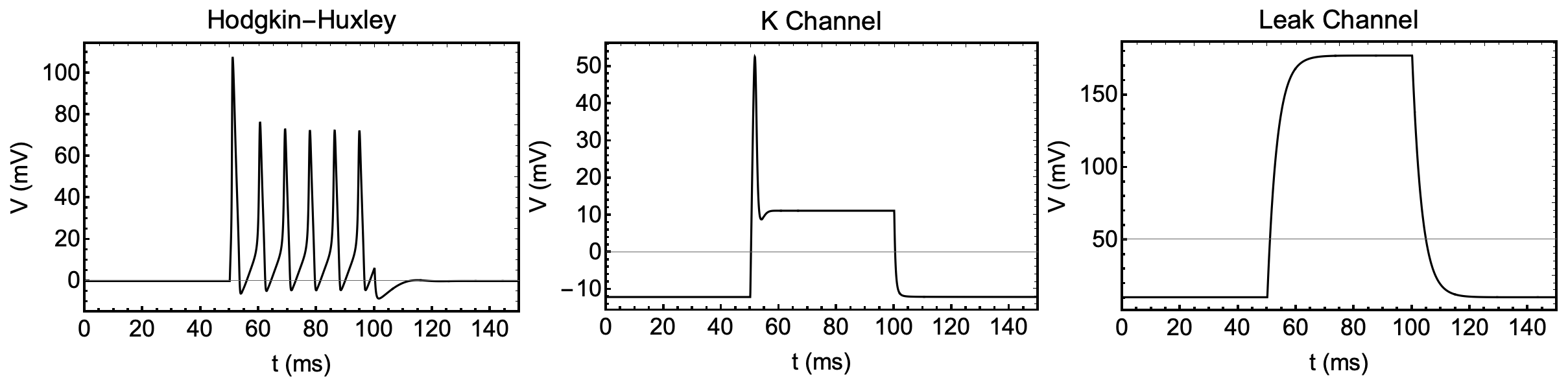}
  \caption[Response of a neuronal membrane with different types of ionic channels to electric stimulus.]{Response of a neuronal membrane with different types and combinations of ionic channels to an electric stimulus I=50 $\mu$A/cm$^2$. The current input was initiated at $t=50$ ms and lasted 50 ms. Each subfigure corresponds to a membrane with different ionic channels, thus being described by characteristic electrophysiological models: the Hodgkin-Huxley model on the left; a model with K channels only in the middle, and a model with Leak channels only on the right.}
  \label{fig:channels_hh_k_l}
\end{figure}

	Figure \ref{fig:channels_na} depicts the results obtained, where the HH's bifurcation diagram was represented in red as a reference. The way each distinctive region of the bifurcation diagram was represented with different types of lines was exactly the same as used in Chapter 2 -- solid lines mark stable points/LC, while dashed lines correspond to unstable ones, a style that will be kept throughout the present chapter. In Figure \ref{fig:channels_hh_k_l}, we compare the membrane potential profile when subjected to I=50 $\mu$A/cm$^2$ for three distinct models.

	A membrane of a neurone that presents only Na channels is mostly unstable for the range of electric stimuli considered. On the other hand, the fixed points are always stable in the K and the Leak models, regardless of the intensity of the external stimulus. Still, regarding the K model, in particular, we see that the position of these fixed points is only slightly away from the HH model's, despite lacking an unstable branch for intermediate intensity values of $I$. Nevertheless, Figure \ref{fig:channels_hh_k_l} demonstrates that the neuronal membrane can still undergo an action potential before reaching a new equilibrium point even in the presence of K channels alone.
	
	To investigate whether the instability in the HH model is brought by the Na's channel, while the position of the fixed points is mainly due to the Potassium channels, alternative models with combinations of two-channel types were put into the test: a K+Na model, a Na+Leak model, and a K+Leak model.

\begin{figure}[!htb]
  \centering
  \includegraphics[width=0.6\textwidth]{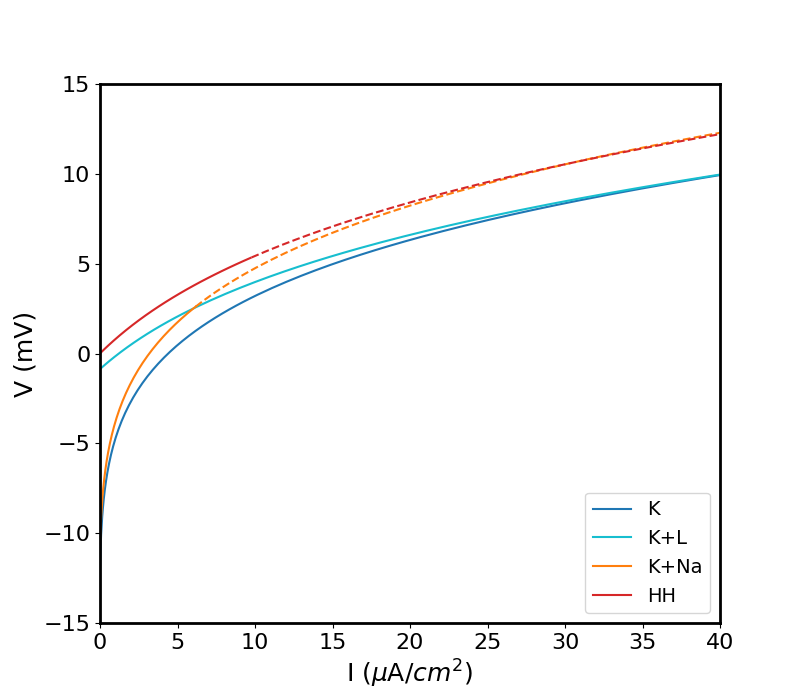}
  \caption[Location and stability of fixed points in electrophysiological models with a different number and combination of ionic channels.]{Location and stability of fixed points in electrophysiological models with a different number and combination of ionic channels. The Figure does not display the fixed points of the Na+Leak model since they were found to be significantly distant from the bifurcation diagram of the Hodgkin-Huxley model.}
  \label{fig:k_kl_kna_hh_bd}
\end{figure}

\begin{figure}[!htb]
  \centering
  \includegraphics[width=\textwidth]{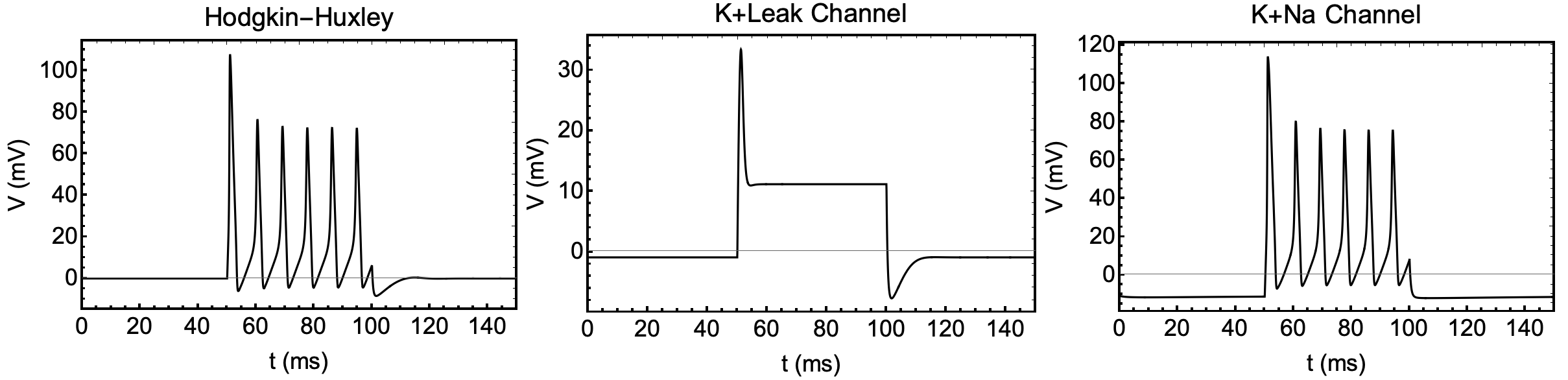}
  \caption[Response of a neuronal membrane with different types and combinations of ionic channels to electric stimulus.]{Response of a neuronal membrane with different types and combinations of ionic channels to an electric stimulus I=50 $\mu$A/cm$^2$. The current input was initiated at $t=50$ ms and lasted 50 ms. Each subfigure corresponds to a membrane with different combinations of ionic channels, thus being described by characteristic electrophysiological models: on the left, the HH's model; in the middle, a model with a combination of K and Leak channels; and, on the right, a model with a combination of Na and K channels.}
  \label{fig:k_kl_kna_hh}
\end{figure}

	After analysing the membrane potential profile of a neurone with both Leak and K channels under various stimulus intensities, our findings indicate that this type of neurone does not exhibit an intermittent or a spiking response in any tested condition. Nonetheless, Figure \ref{fig:k_kl_kna_hh} b) shows that a neurone with a combination of these channels can still generate an action potential.
		
	On the other hand, the results obtained for the Na+K model, exemplified by Figures \ref{fig:k_kl_kna_hh} and \ref{fig:k_kl_kna_hh_bd} c), provide clear evidence that the essential characteristics of the HH model remain intact when the Leak channel is excluded, including the intermittent membrane potential response and spiking behaviour within a specific intensity range of electric stimuli.
	
	Based on these results, three conclusions can be drawn. Firstly, the contribution of the Leak channel to the membrane potential evolution is not dominant and can be disregarded, particularly for intermediate to high-intensity stimuli. Secondly, the K channel plays a crucial role in determining the location of fixed points in the bifurcation diagram $V^{*}(I)$, highlighting its significance in shaping the overall dynamics of the system. Finally, the Na channel contributes to the emergence of unstable membrane responses for stimuli of intermediate intensity, originating, together with the K channel, two Hopf bifurcation points in the $V^{*}(I)$ profile and phenomena of intermittency and spiking activity. Based on these insights, it is possible to simplify the Hodgkin-Huxley model by disregarding two parameters: $V_L$ and $g_L$. 

\subsection{Connections between Gating Variables and Model Reduction}

To complement the analysis of the membrane potential's evolution under several electric stimuli, we investigated the dynamics of the gating variables $n$, $m$, and $h$. During our investigation, a consistent relationship between $n$ and $h$ emerged while examining the profiles of $n(t)$, $m(t)$ and $h(t)$ for different stimulus intensities.

\begin{figure}[!htb]
  \begin{subfigmatrix}{2}
    \subfigure[Zoomed-out view]{\includegraphics[width=0.49\linewidth]{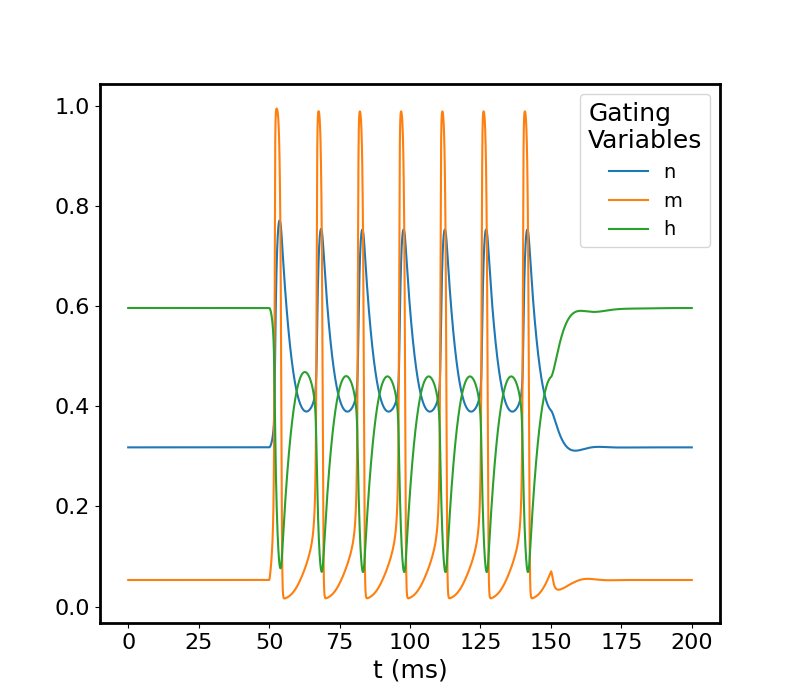}}
    \subfigure[Zoomed-in view]{\includegraphics[width=0.49\linewidth]{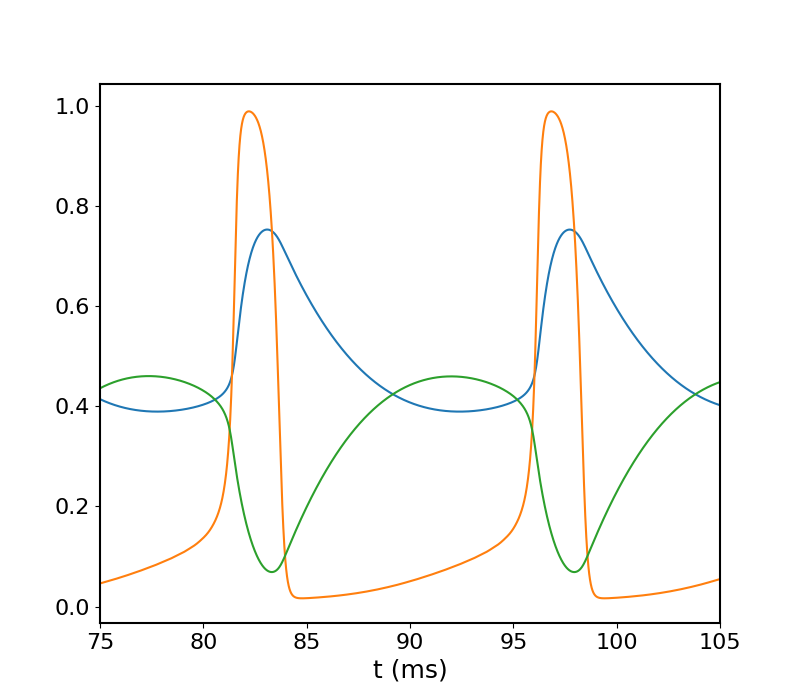}}
  \end{subfigmatrix}
  \caption[Gating variables' time response in the Hodgkin-Huxley model.]{Gating variables' time response when subjected to a stimulus $I=10$ $\mu$A/cm$^2$ that is turned on from $t=50$ ms and lasts 100 ms.}
  \label{fig:nmh}
\end{figure}

\begin{figure}[!htb]
  \begin{subfigmatrix}{2}
    \subfigure[Zoomed-in view]{\includegraphics[width=0.6\linewidth]{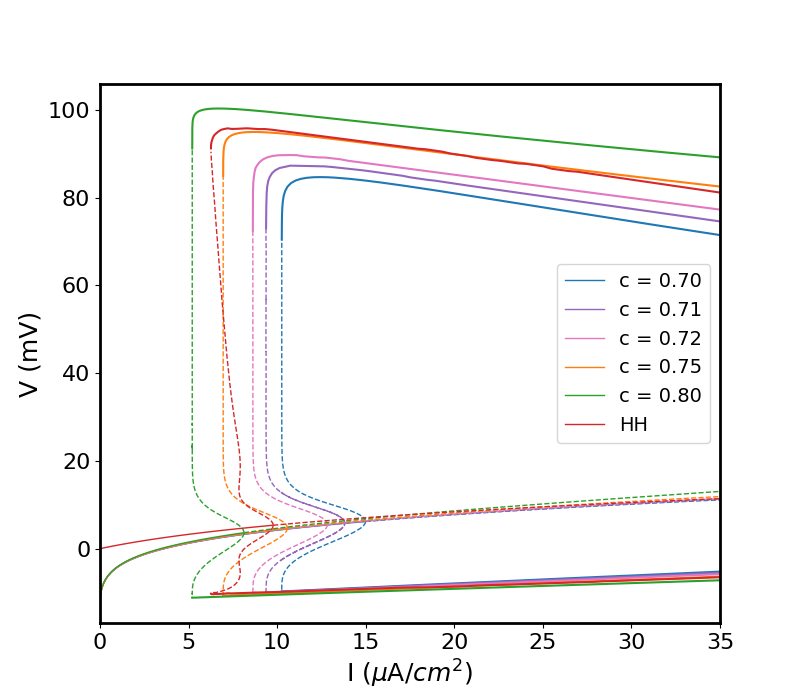}}
    \subfigure[Zoomed-out view]{\includegraphics[width=0.6\linewidth]{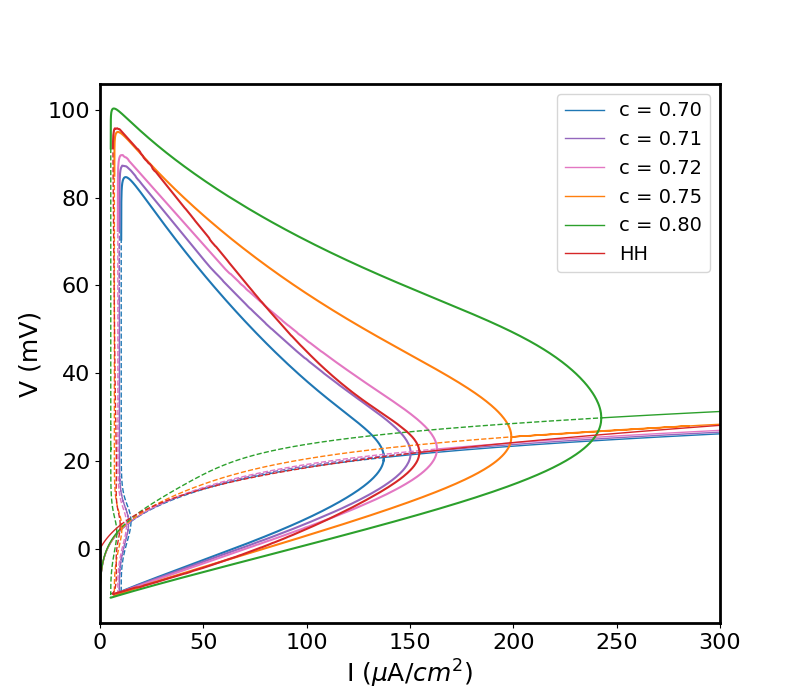}}
  \end{subfigmatrix}
  \caption[Comparison between the bifurcation diagram of reduced models with three independent variables and the Hodgkin-Huxley's model.]{Comparison between the bifurcation diagram of reduced models with three independent variables and the Hodgkin-Huxley's model. In these reduced models, which are described by the system of equations \ref{eq:3dmodel}, the gating variable $h$ of the HH model was replaced by $c-n$, and several values of the constant $c$ were put into test, as presented in the Figure.}
  \label{fig:c}
\end{figure}

\begin{figure}[!htb]
  \centering
  \includegraphics[width=\textwidth]{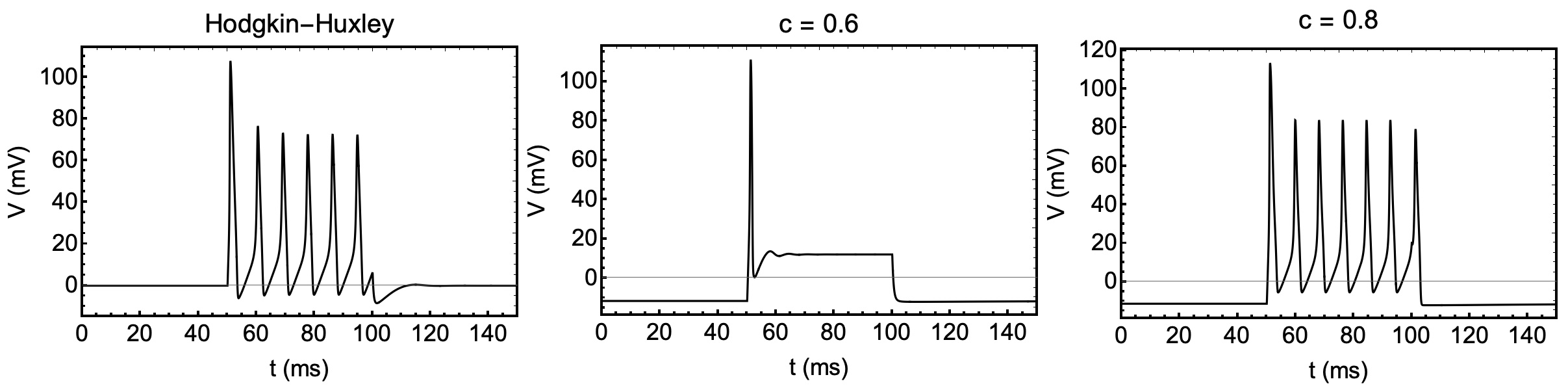}
  \caption[Membrane's response to electric stimulus considering three-dimensional models.]{Membrane's potential response to an external stimulus $I = 50$ $\mu$A/cm$^2$ that is activated at $t=50$ ms and lasts 50 ms. Different electrophysiological models were considered. The membrane's response using the HH model is depicted on the left. On the central and right images, models governed by the system of equations \ref{eq:3dmodel} were put into test, using different values for the constant $c$.}
  \label{fig:cAP}
\end{figure}

As it becomes clear in Figure \ref{fig:nmh}, the gating variable $h$ appears to mirror $n$, which suggests a correlation between the conductivity of the Na and the K channels. This correlation, which has also been reported by \citet{fitzhugh1961445}, motivated us to test the model portrayed in equation \ref{eq:3dmodel}. In this model, the gating variable $h$ was replaced by $c-n$, and various values of $c$ were tested by analysing the bifurcation diagram of this reduced model and the membrane potential response to different electric stimuli. The results are presented in Figures \ref{fig:c} and \ref{fig:cAP}.

\begin{subequations}
\begin{equation}
C_m\frac{dV}{dt} = I - g_Kn^4(V-V_K) - g_{Na}m^3(c-n)(V-V_{Na})
\end{equation}\label{eq:3dmodel}
\vspace{-2ex}
\begin{equation}
\frac{dm}{dt} = \alpha_m(V)(1-m)-\beta_m(V)m
\end{equation}
\vspace{-2ex}
\begin{equation}
\frac{dn}{dt} = \alpha_n(V)(1-n)-\beta_n(V)n
\end{equation}
\end{subequations}

 The striking similarities to the HH bifurcation diagram in terms of global characteristics of these reduced models suggest that the Na channel's behaviour can be described by a combination of the gating variables $m$ and $n$, obviating the need for an additional independent variable $h$. Particularly, for currents below approximately 50 $\mu$A/cm$^2$, the bifurcation line corresponding to the model with $c=0.75$ closely aligns with the HH diagram, exhibiting only minor discrepancies in the minimum and maximum values of periodic cycles. When $c\approx 0.70$, the coordinates of the fixed points, both stable and unstable, closely coincide, further supporting the notion of a direct relationship between Na and K channels. While \citet{fitzhugh1961445} considered $c=0.85$, our analysis reveals that the closest approximation to the HH bifurcation diagram is achieved when $c=0.71$. Hence, we replaced $h$ by $c-n=0.71-n$ in subsequent tests and simulations, which preserved the important biological features of the HH model while reducing its complexity and facilitating the exploration of the system's behaviour.

We now investigate the impact of approximating the gating variables $n$ and $m$ by their steady-state values on the system's behaviour. Due to the rapid response of the gating variable $m$, we anticipated that approximating it by its steady-state value would yield satisfactory results. However, we conducted separate tests for the two gating variables to validate this assumption. Additionally, we performed a combined approximation where only the membrane potential $V$ remained as the sole dynamical variable.

\begin{figure}[!htb]
  \begin{subfigmatrix}{2}
    \subfigure[Zoomed-in view]{\includegraphics[width=0.49\linewidth]{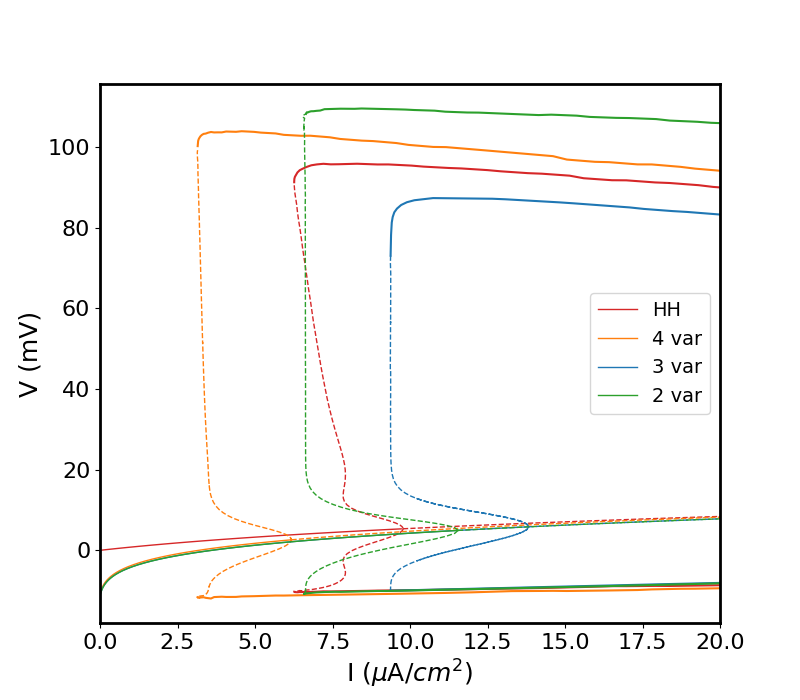}}
    \subfigure[Zoomed-out view]{\includegraphics[width=0.49\linewidth]{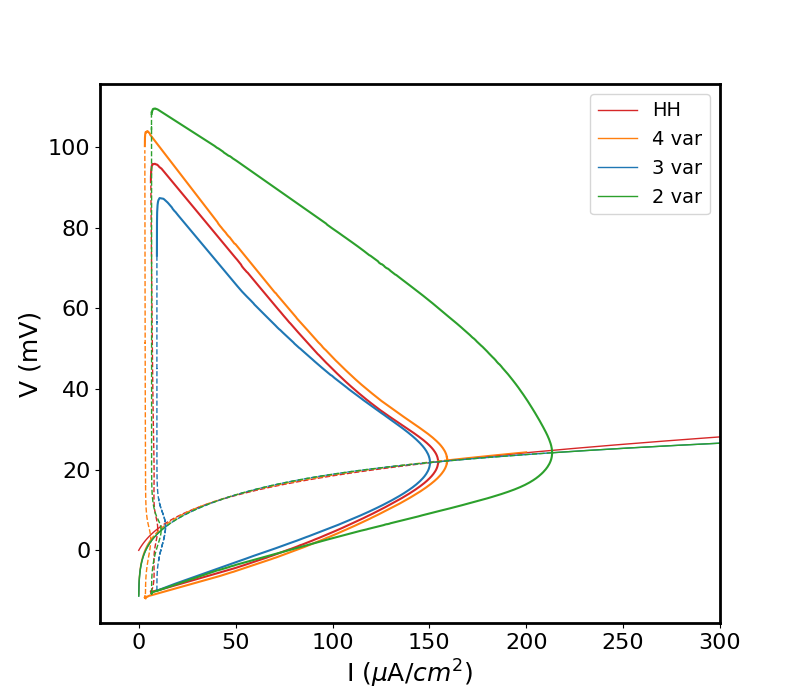}}
  \end{subfigmatrix}
  \caption[Comparison between bifurcation diagrams of reduced models with varying numbers of independent variables and the Hodgkin-Huxley model.]{Comparison between the bifurcation diagrams of different reduced models and the Hodgkin-Huxley model. The HH model's bifurcation diagram is depicted in red as a reference. The orange diagram represents a model without the leak channel, still utilising four variables but excluding two parameters from the HH model. In blue, we illustrate the bifurcation diagram of an electrophysiological model where the variable $h$ is replaced by $0.71-n$, and the leak channel is disregarded, resulting in the three-variables system (\ref{eq:3dmodel}) with $c=0.71$. Lastly, the green diagram showcases the bifurcation diagram of a model where the gating variable $m$ is replaced by its asymptotic value $m_{\infty}$ and $h$ is approximated by $0.71-n$, resulting in the two-variables system \ref{eq:2dmodel} with $c=0.71$.}
  \label{fig:hh_nak_c_2v}
\end{figure}

The results presented in Figure \ref{fig:hh_nak_c_2v} show that approximating $m$ by its steady-state value $m_{\infty}$ yielded reasonable outcomes, with all the global characteristics of the resulting bifurcation diagram matching those of the HH model. For this model, the first Hopf bifurcation occurs at $I = 11.5478$ $\mu$A/cm$^2$ and the second at $I=213.352$ $\mu$A/cm$^2$. On the other hand, approximating $n$ by its steady-state value $n_{\infty}$ led to a model without Hopf bifurcations and periodic cycles and where all the fixed points were stable.

By incorporating $m_{\infty}$ into the system of equations, we achieved the objective of reducing the HH model from four to two independent variables - equation \ref{eq:2dmodel}. Furthermore, neglecting the influence of the Leak channel eliminated two non-essential parameters that had minimal impact on the system's behaviour.

\begin{subequations}
\begin{equation}
C_m\frac{dV}{dt} = I - g_Kn^4(V-V_K) - g_{Na}m_{\infty}(V)^3(c-n)(V-V_{Na})
\end{equation}\label{eq:2dmodel}

\begin{equation}
m_{\infty}(V) = \frac{\alpha_m(V)}{\alpha_m(V)+\beta_m(V)}
\end{equation}

\begin{equation}
\frac{dn}{dt} = \alpha_n(V)(1-n)-\beta_n(V)n
\end{equation}
\end{subequations}

\begin{figure}[!htb]
  \centering
  \includegraphics[width=0.99\textwidth]{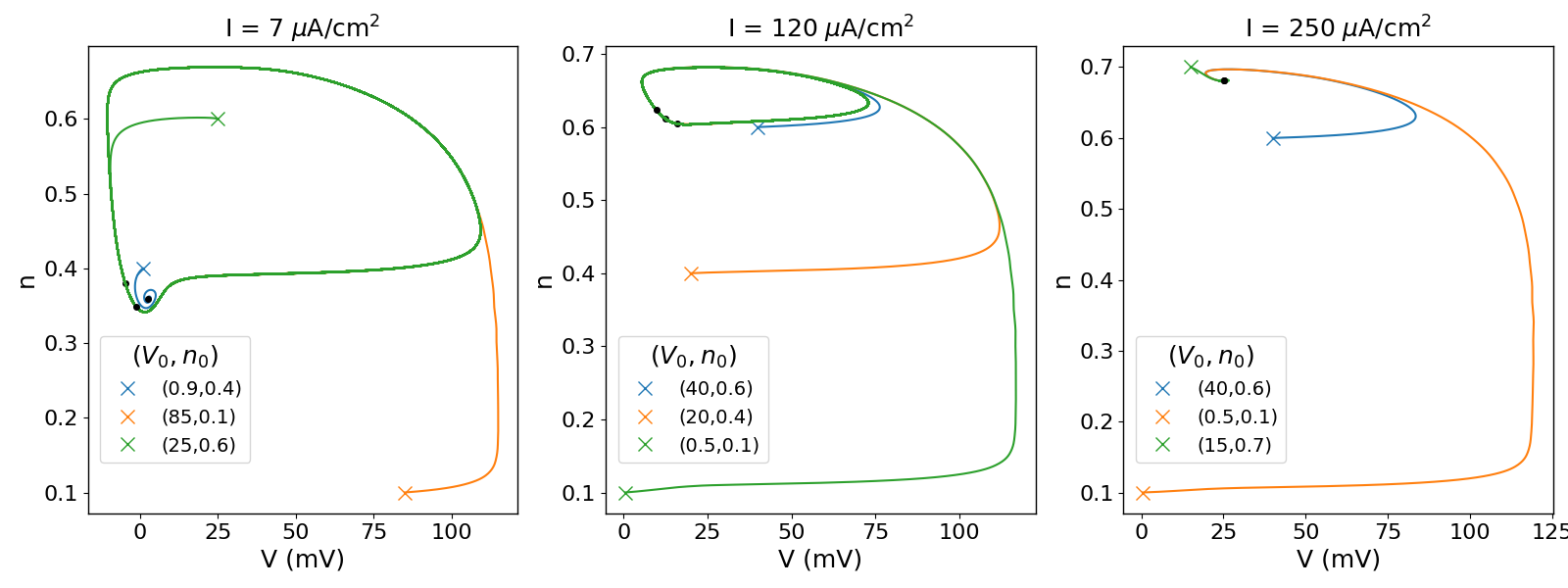}
  \caption[Phase space of the bi-dimensional reduced model.]{Phase space of the bidimensional model ruled by the system of equations \ref{eq:2dmodel} with $c=0.71$. We considered three different external stimuli $I$ and distinct initial conditions, marked with crosses and labelled on each subfigure. All the simulations were run until $t=1000$ ms, and the final point for each initial condition and electric stimulus was marked with a black dot.}
  \label{fig:phase}
\end{figure}

The reduced model not only accelerates computational simulations but also enhances the interpretability and comprehension of the model's behaviour. To exemplify, in the Figure \ref{fig:phase}, we depict the membrane potential and the gating variable $n$ evolution for 1 s, starting at different initial conditions $(V_0, n_0)$ and for three different stimuli.

When $I$=7 $\mu$A/cm$^2$, the membrane potential can exhibit intermittent dynamics despite being in a stable regime. That becomes clear by looking at the left image of Figure \ref{fig:phase} where, for two different initial conditions, the membrane potential presents several consecutive action potentials, ending up in distinct final points after 1 s. However, the membrane potential and gating variable $n$ exhibit rapid decaying oscillations for other initial conditions, converging to a stable fixed point. This behaviour is verified, for example, for $(V_0, n_0) = (0.9, 0.4)$, when $I$=7 $\mu$A/cm$^2$, as it can be observed in the same Figure.

As the injected current increases, the system transitions to a region of the bifurcation diagram where the fixed points become unstable, and a limit cycle arises. This corresponds to the emergence of sustained rhythmic spiking activity independent of the specific initial conditions $(V_0, n_0)$. Regardless of the starting values for $n$ and $V$, the system converges to the same limit cycle trajectory in the phase space, as seen in the central image of Figure \ref{fig:phase}. 

Beyond the second bifurcation point, the membrane potential no longer exhibits spiking activity. Instead, as we exemplify for $I$ = 250 $\mu$A/cm$^2$ in the right image of Figure \ref{fig:phase}, even for initial conditions that are significantly distant from the respective fixed point, the system can eventually undergo an initial action potential but quickly converges to the final equilibrium state.

\section{Influence of the System's Parameters on Spiking Behaviour}

This section explores the intrinsic spiking behaviour of the two-variable model developed earlier. We examine the effect of the electric stimulus intensity on the membrane potential's spiking frequency, denoted $f_s$. Subsequently, we investigate the influence of several model parameters, such as the ionic conductivities $g_K$ and $g_{Na}$, along with the corresponding Nernst potentials $V_K$ and $V_{Na}$. This enables us to predict the response that cells with slightly different characteristics present when subjected to electric stimuli and better understand the distinct functions played by each ion channel in this phenomenon. 

It is worth noting that this comparison was only feasible because the spiking phenomenon, regardless of the values of the parameters tested, always remained regular -- in frequency  and in amplitude -- for a given combination of parameters and stimulus intensity $I$.

\begin{figure}[!htb]
  \centering
  \includegraphics[width=0.9\textwidth]{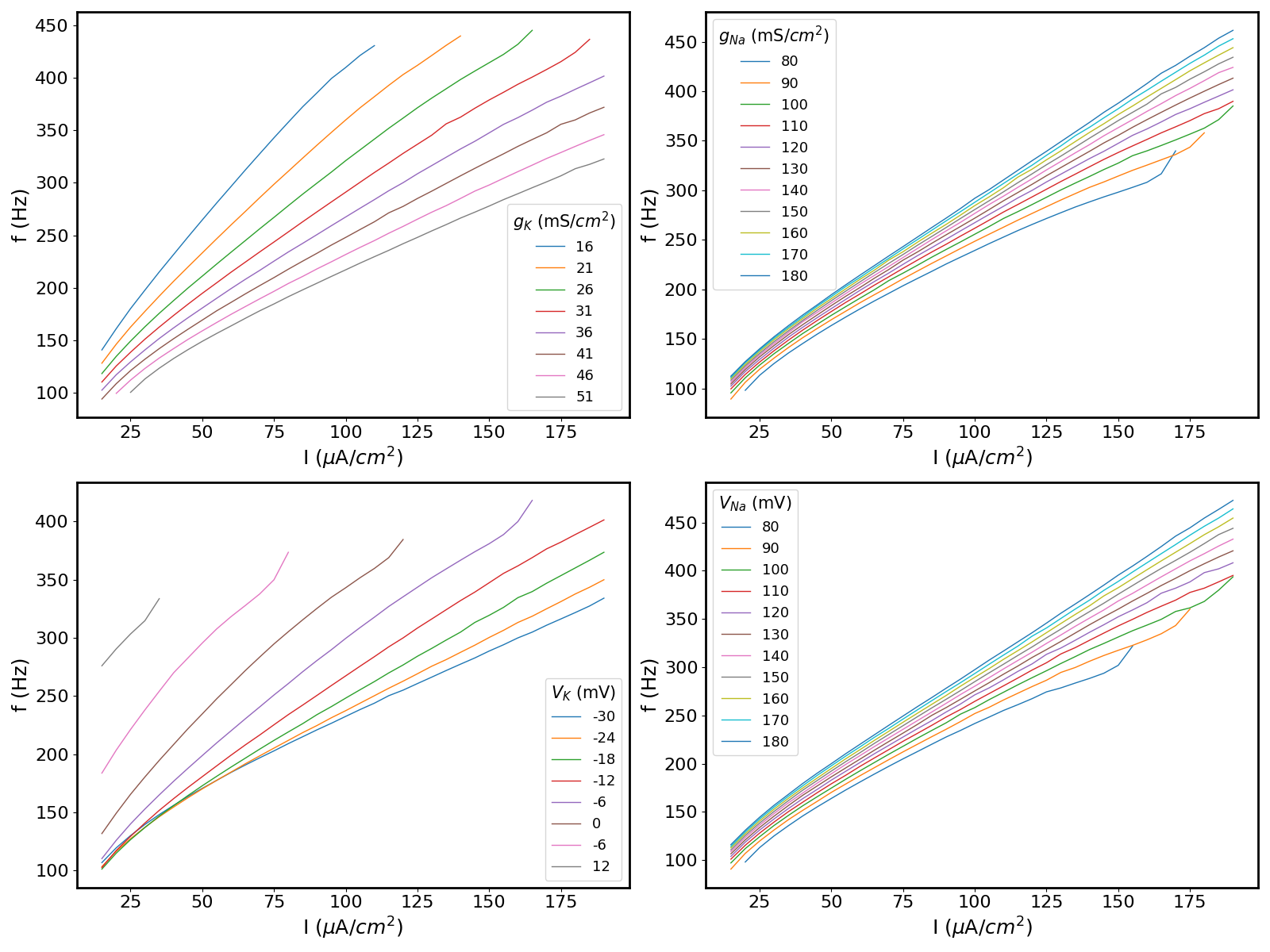}
  \caption[Dependence of the spiking frequency on the electric stimuli and on the electrophysiological model's conductivities and Nerst potentials.]{Dependence of the spiking frequency on the electric stimuli $I$ and on the system's parameters $g_{Na}$, $g_K$, $E_K$ and $V_{Na}$. While each parameter was varied, the others were kept constant at the original model's values ($g_{Na}=120$ mS/cm$^2$, $g_K = 36$ mS/cm$^2$, $V_K = -12$ mV and $V_{Na} = 115$ mV). We considered the electrophysiological model governed the system of equations \ref{eq:2dmodel} with $c=0.71$.}
  \label{fig:f(i)}
\end{figure}

Based on the results obtained and presented in Figure \ref{fig:f(i)} it is evident that $f_s$ increases with the stimulus' intensity $I$, irrespective of the modifications made in the model's parameters. Moreover, the spiking behaviour of the system is significantly affected by the conductivities and ionic Nernst potentials. Specifically, $f_s$ increases as $g_K$ decreases for a given stimulus, indicating that higher Potassium conductance allows for a more frequent spiking. Conversely, increasing $g_{Na}$ leads to a consistent increase in the spiking frequency. Increasing either $V_{Na}$ or $V_{K}$ results in an increase in the spiking frequency, although changes in $V_{K}$ have a more pronounced impact on the neuronal behaviour than $V_{Na}$.

\begin{figure}[!htb]
  \begin{subfigmatrix}{2}
    \subfigure[Bifurcation diagram of systems with different conductivities $g_K$]{\includegraphics[width=0.49\linewidth]{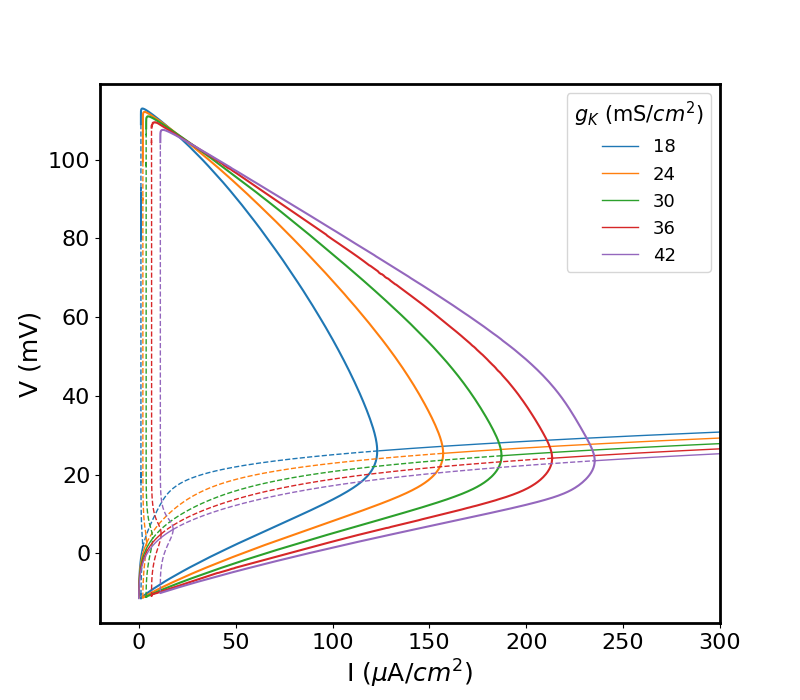}}
    \subfigure[Bifurcation diagram of systems with different Nerst potentials $V_K$.]{\includegraphics[width=0.49\linewidth]{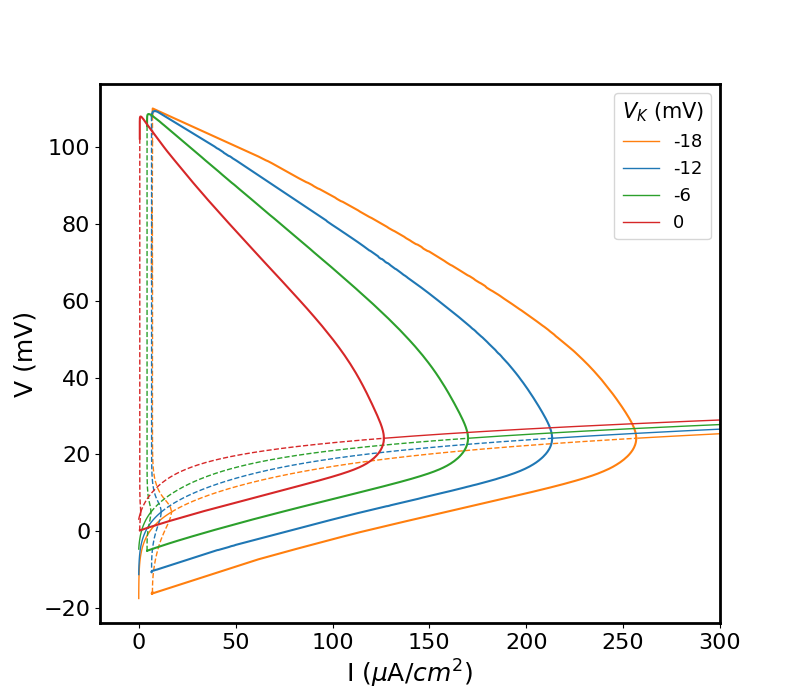}}
  \end{subfigmatrix}
  \caption[Influence of the Potassium Channel's parameters in the system's bifurcation diagram.]{Influence of $V_K$ and $g_K$ in the bifurcation diagram of the electrophysiological model governed by the system of equations \ref{eq:2dmodel} with $c=0.71$.}
  \label{fig:gk_ek_bd}
\end{figure}

The overall effect of these parameters in the systems' bifurcation diagrams is also non-negligible, as shown in the Figure \ref{fig:gk_ek_bd} for different values of $g_K$ and $V_K$.

\begin{figure}[!htb]
  \begin{subfigmatrix}{2}
    \subfigure[Spiking frequency of membranes with different capacitances.]{\includegraphics[width=0.49\linewidth]{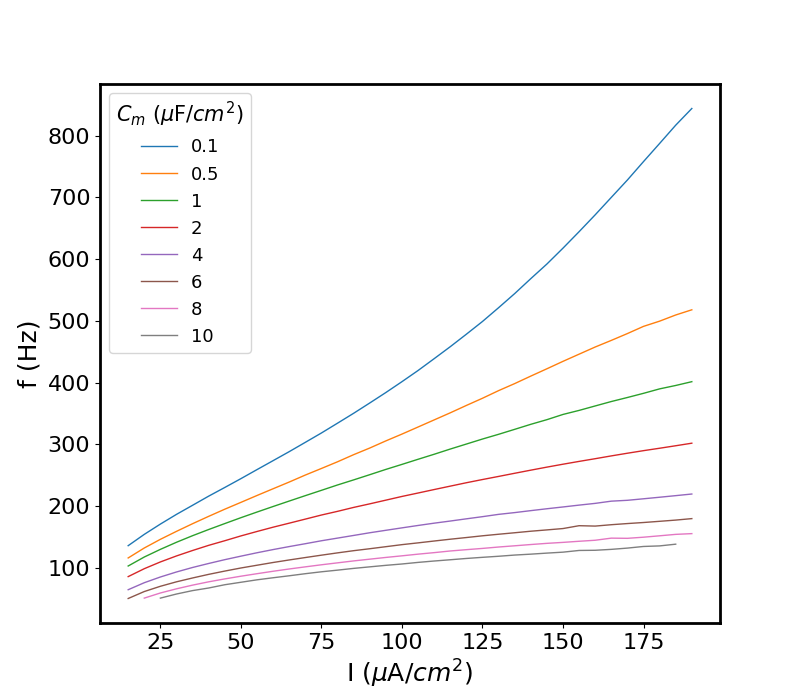}}
    \subfigure[Bifurcation diagram associated with membranes with different capacitances.]{\includegraphics[width=0.49\linewidth]{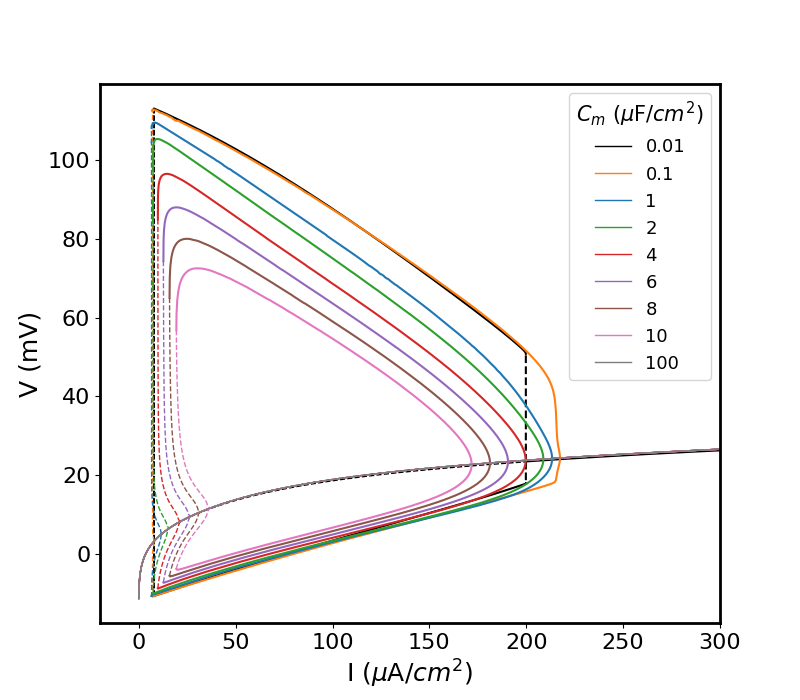}}
  \end{subfigmatrix}
  \caption[Influence of the cell capacitance in the electrophysiological model's features, including the periodic cycle's features, the location of the Hopf bifurcation points and the spiking frequency.]{Influence of $C_m$ in the features of the electrophysiological model's ruled by equation \ref{eq:2dmodel} with $c=0.71$, including the periodic cycle's features, the location of the Hopf bifurcation points and the spiking frequency.}
  \label{fig:cm}
\end{figure}

Another parameter of the system that significantly influences the spiking frequency and its bifurcation diagram is the capacitance $C_m$, which, so far, we have always considered to be 1 $\mu$F/cm$^2$. Just like we did for other parameters of our reduced model, we evaluated $f_s(I)$ for several cellular capacitances, as depicted in Figure \ref{fig:cm} a). The associated bifurcation diagrams are depicted in Figure \ref{fig:cm} b).

Our findings revealed that, as the capacitance of the cell increases, the spiking frequency decreases. This can be explained by the fact that the cell capacitance represents the ability of the neuronal membrane to store and release charge: a higher capacitance means that more charge is required to change the membrane potential, which, as a result, implies that it takes longer for the membrane potential to reach the threshold level required for an action potential to occur. In extreme cases, as it was verified for $C_m =$ 100 $\mu$F/cm$^2$, all the fixed points in the range of stimuli $I\in\:[0,300]$ $\mu$A/cm$^2$ are stable, and there are no Hopf bifurcation points.

On the other hand, our results suggest that a cell with a small ability to store charge is more likely to respond with a train of high-frequency spikes, even for small-intensity stimuli. Additionally, Figure \ref{fig:cm} b) shows that decreasing the cell capacitance results in higher voltage peaks during spiking activity. 

For a cell capacitance of $C_m = 0.1 \mu$F/cm$^2$, there is a noticeable change in the shape of the bifurcation diagram near the second Hopf bifurcation point. However, this change becomes even more prominent for $C_m = 0.01 \mu$F/cm$^2$, in which case, the limit cycle near that bifurcation point is unstable, resulting in a completely distinct response by the cell membrane.

\begin{figure}[!htb]
  \begin{subfigmatrix}{2}
    \subfigure[Zoom-out view]{\includegraphics[width=0.49\linewidth]{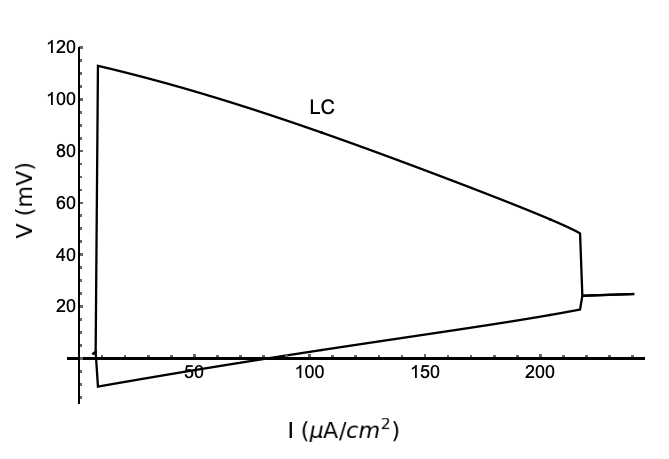}}
    \subfigure[Zoom-in view]{\includegraphics[width=0.49\linewidth]{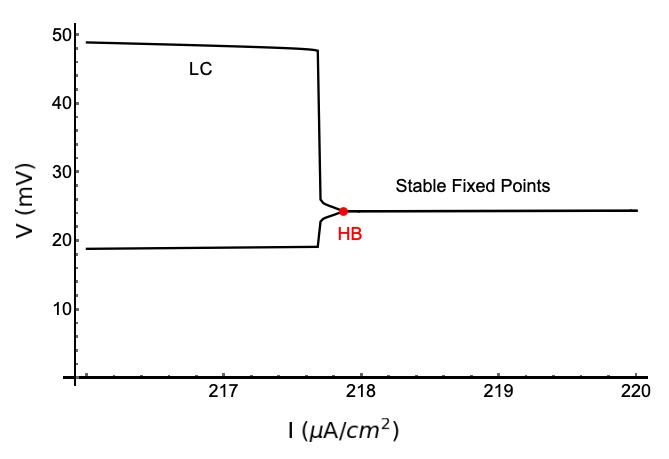}}
  \end{subfigmatrix}
  \caption[Limit cycles of the electrophysiological model with a small cell capacitance.]{Limit cycles of the reduced model for $Cm =$ 0.01 $\mu$F/cm$^2$ with the second Hopf bifurcation being represented by a red dot.}
  \label{fig:cm-lc}
\end{figure}

Given the significance of this change in the system's bifurcation diagram, we conducted a more detailed analysis focusing on the system's behaviour when $C_m = 0.01 \mu$F/cm$^2$. In Figure \ref{fig:cm-lc}, we depict the shape of the limit cycle (without considering its stability). The diagram reveals a sharp reduction in the width of the limit cycle solutions near the second Hopf bifurcation point. This indicates an extremely high sensitivity of the membrane to small changes in the electric stimulus $I$.

\begin{figure}[!htb]
  \centering
  \includegraphics[width=0.99\textwidth]{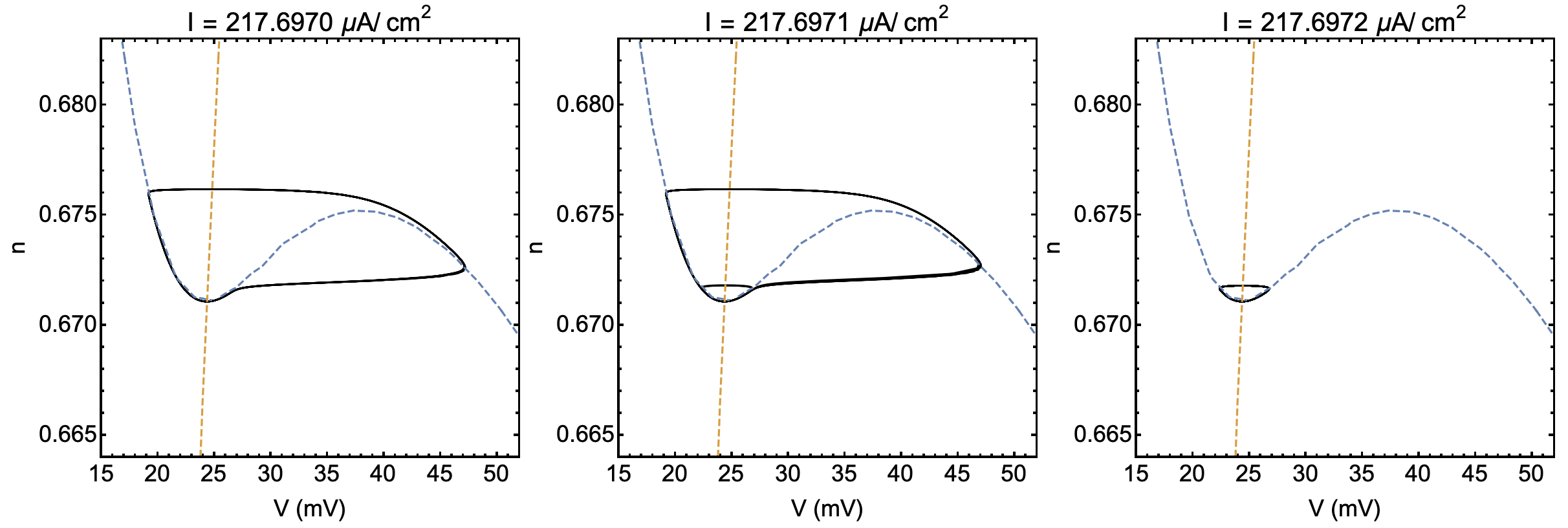}
  \caption[Canards in neurones with a small capacitance.]{Phase space trajectories when the membrane is subjected to three slightly distinct stimuli near the second Hopf bifurcation of the system of equations \ref{eq:2dmodel} with $c=0.71$ and $C_m=$ 0.01 $\mu$F/cm$^2$. The blue dashed line is the nullcline of the membrane potential $V$, while the yellow dashed line corresponds to the nullcline of the gating variable $n$. The fixed point is at the intersection of the nullclines. We considered that the membrane evolved from its resting state. The plots represent the phase space trajectories in the interval $t\in$ [30,40] ms after each stimulus had been initiated. The image on the left corresponds to a canard with a head. The central image depicts the phase space trajectory during the transition from a canard with a head to a small canard, as depicted in the right image.}
  \label{fig:canards}
\end{figure}

\begin{figure}[!htb]
  \centering
  \includegraphics[width=0.99\textwidth]{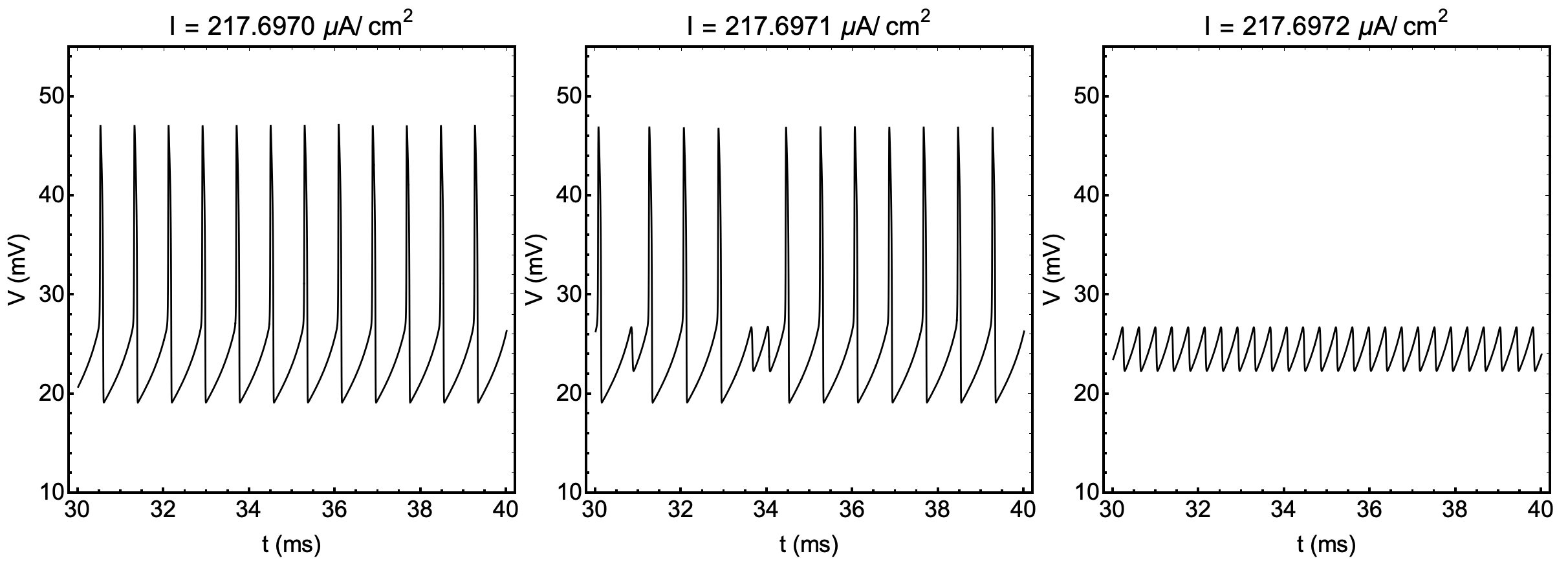}
  \caption[Membrane potential's response nearby the second Hopf bifurcation point when the neurone has a small capacitance.]{Different types of response presented by the membrane's potential when subjected to three slightly distinct stimuli nearby the second Hopf bifurcation of the system of equations \ref{eq:2dmodel}, with $c=0.71$ and $C_m=$ 0.01 $\mu$F/cm$^2$. We once again considered that, at $t=0$, the cell membrane started in its resting state. The representation of $V(t)$ is restricted to the interval $t\in$ [30,40] ms for more enhanced visualisation of the cell+s response.}
  \label{fig:canards-V(t)}
\end{figure}

As depicted in Figure \ref{fig:canards}, we observed the existence of different types of canards within a very narrow region of the parameter space: a canard with a head and a small canard. As shown in Figure \ref{fig:canards-V(t)}, even a small change in the electric stimulus $I$ in the order of $10^{-4}$ $\mu$A/cm$^2$ is sufficient to generate completely different membrane potential responses. 

\newpage

\section{Signal Propagation along the Axon}

In this section, we simulate the propagation of electrical signals along the axon. To achieve this, we adopted a spatially extended version of the previous model, where the axon was discretised into multiple ($N$) nodes and where the potential along the axon's length depends only on the length variable, $x$, and not on radial or angular variables. Each pair of nodes defines the starting and ending point of a small myelinated axon segment with length $dx$. In this work, we have considered $dx=1$, corresponding to a dimensionless measure of the respective quantity. In experimental tests, the value of $dx$ should be changed to its real value, depending on the neuronal cell under testing. This compartmental approach is also used by the simulation packages NEURON \cite{neuron} and GENESIS \cite{Bower1994TheBO}, but differs from other approaches (for Ermentrout).
(see the discussion below).

\begin{figure}[!htb]
  \centering
  \includegraphics[width=0.9\textwidth]{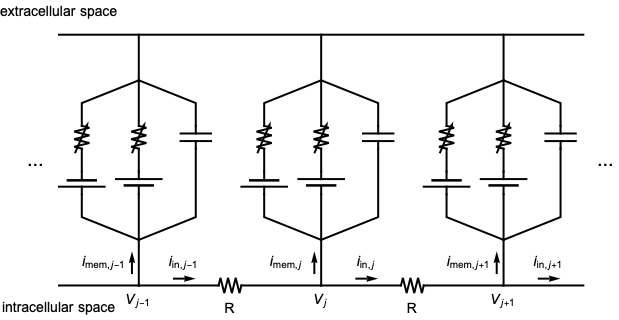}
  \caption[Electric scheme of the compartmental model.]{Electric scheme of the compartmental model. $I_{j-1,j}$ corresponds to the intracellular current that goes from the node with potential $V_{j-1}$ to the node with potential $V_j$; $I_{ion,j}$ is the current originated by the passage of ions through the ion-specific channels in the compartment $j$; $I_{mem,j}$ is the membrane current in the compartment $j$, which is a combination of $I_{ion, j}$ and the current of the capacitor.}
  \label{fig:spatial_propagation/circuit}
\end{figure}

To account for the intracellular resistance, we introduce a resistance parameter $R$ that represents the resistance between neighbouring compartments in the intracellular space. We assumed the extracellular space resistance to be negligible, implying that the potential outside the cell remains constant. In this model, the electric stimulus is transmitted from the soma to the first node of the axon ($N=1$ and $x=0$). Figure \ref{fig:spatial_propagation/circuit} schematically illustrates three nodes of this compartmental model.

Looking at Figure \ref{fig:spatial_propagation/circuit} and using Kirchhoff laws, our compartmental model can be compiled in the following equations

\begin{subequations}
\begin{equation}
	\begin{aligned}
	&C_m \frac{dV_1}{dt} = I - \bigg(\frac{V_1-V_2}{R}\bigg) - I_{ion}(V_1, n_1)\\
	C_m \frac{dV_j}{dt} = &\bigg(\frac{V_{j-1}-V_{j}}{R}\bigg) - \bigg(\frac{V_{j}-V_{j+1}}{R}\bigg) - I_{ion}(V_j, n_j), \:\: 1 < j < N \\
	&C_m \frac{dV_N}{dt} = \bigg(\frac{V_{N-1}-V_{N}}{R}\bigg) - I_{ion}(V_N, n_N)
    \end{aligned}
    \label{prop-model}
\end{equation}

\begin{equation}
	I_{ion}(V_j, n_j) = g_Kn_j^4(V_j-V_K) + g_{Na}m_{\infty}(V_j) ^3(c-n_j), \:\: j \in [1,N],
\end{equation}

\begin{equation}
\frac{dn_j}{dt} =\alpha_n(V_j)(1-n_j)-\beta_n(V_j)n_j , \:\: j \in [1,N],
\end{equation}

\begin{equation}
m_{\infty}(V_j) = \frac{\alpha_m(V_j)}{\alpha_m(V_j)+\beta_m(V_j)} , \:\: j \in [1,N]
\end{equation}\label{eq:propagation}
\end{subequations}

In our tests and simulations, and before considering any possible perturbations in the axon's membrane potential, we initialised the nodes in the resting state of $V$ and $n$ in the absence of any external electric stimulus $I$: $V_0 \approx -11.3554$ mV and $n_0 \approx 0.1657$.

\subsubsection{Membrane Potential Evolution along the Axon}

Using the Euler method, we numerically solved the differential equations of the proposed compartmental model for stimuli in the range [1, 300] $\mu$A/cm$^2$. To illustrate the results, we present $V(t)$ at three distinct nodes when $x=0$ is subjected to three different electric stimulus intensities $I$. Each stimulus corresponded to a specific regime in the bifurcation diagram of our reduced model. Furthermore, we considered three distinct resistances $R$ for each stimulus, so that the influence of this parameter could be better described. In this set of tests, we considered an axon with $N=200$ nodes.

\begin{figure}[!htb]
  \centering
  \includegraphics[width=0.99\textwidth]{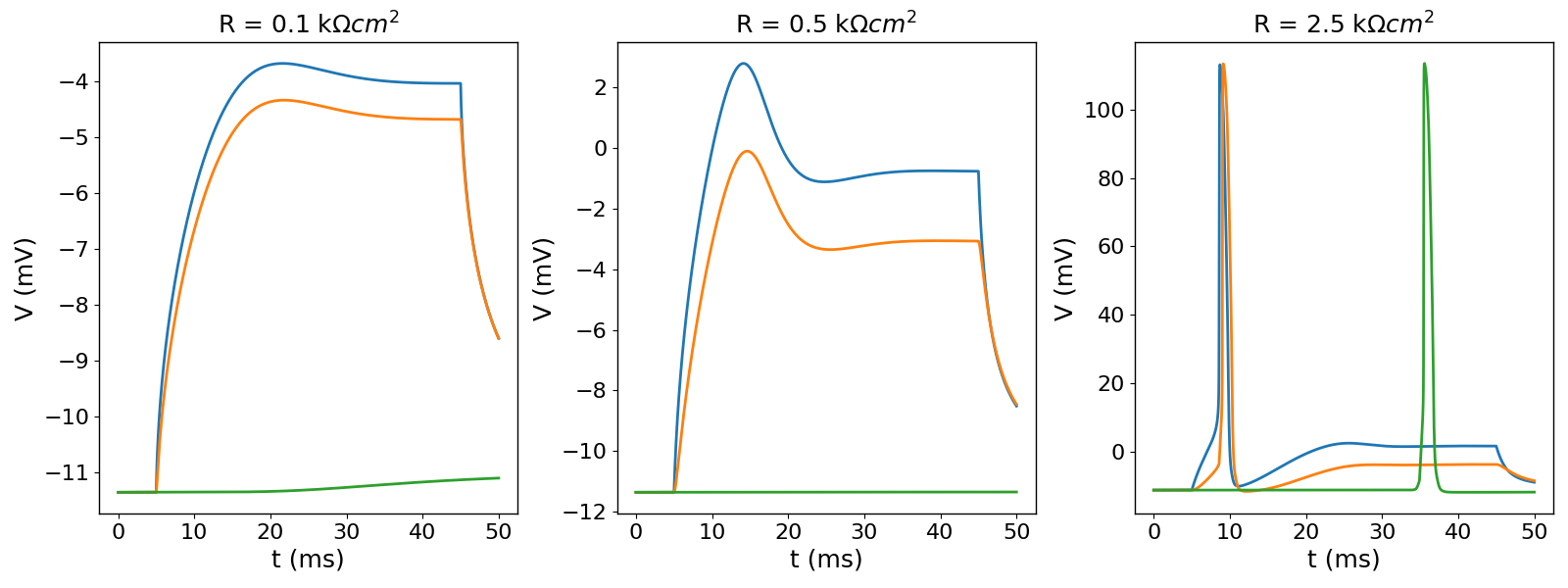}
  \caption[Response of the membrane potential at different sites of the axon when the neuron is subjected to a small intensity electric stimulus.]{Membrane potential's response to an electric stimulus $I=7.5$ $\mu$A/cm$^2$ that is turned on at $t=5$ ms and turned off at $t=45$ ms, for three distinct intracellular resistances. The different colours correspond to different nodes of the axon: $x=0$ in blue, $x=1$ in orange, and $x=50$ in green.}
  \label{fig:AP-7.5}
\end{figure}

As it becomes clear in the Figure \ref{fig:AP-7.5}, the cell membrane tends to converge to its resting state without generating an action potential for lower-intensity stimuli and low intracellular resistances. This behaviour is evident for a resistance value of R = 0.1 $k\Omega$/cm$^2$, where the axon nodes stabilise at a characteristic resting potential. Slightly higher resistances result in minimal changes in most 200 nodes. However, for greater intracellular resistances, an action potential can be initiated at the first node and propagate through the neurone.

\begin{figure}[!htb]
  \centering
  \includegraphics[width=0.99\textwidth]{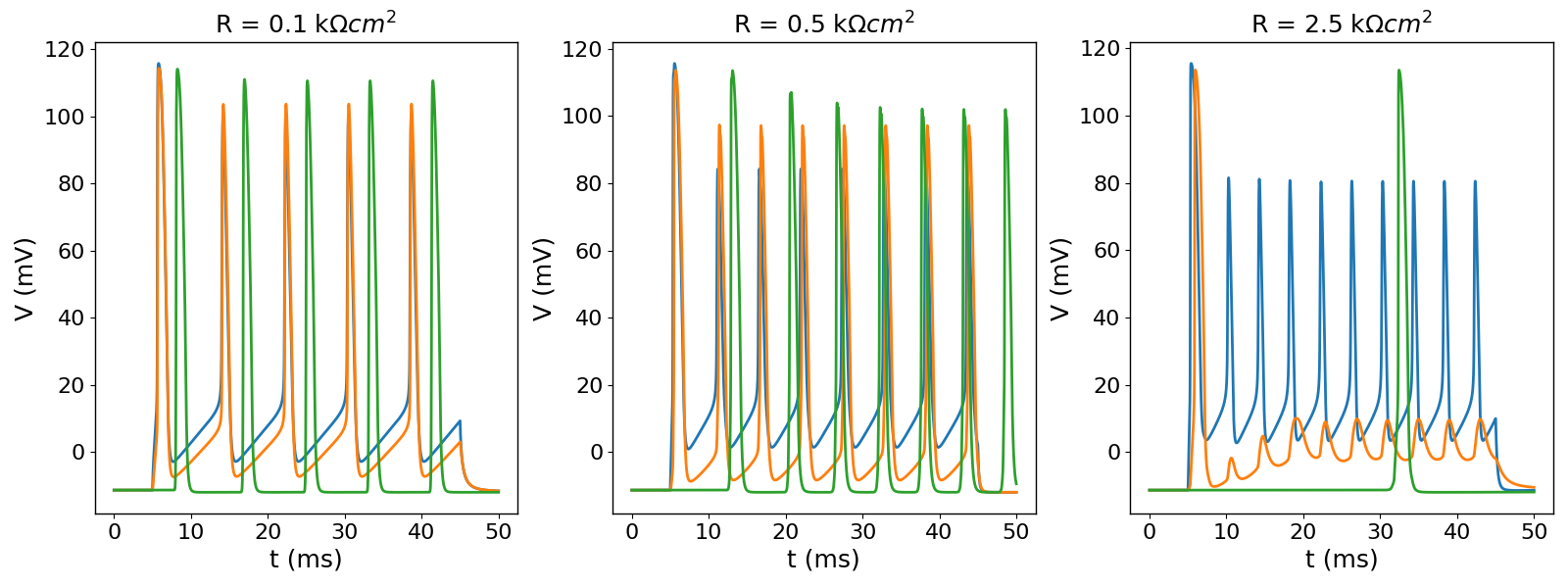}
  \caption[Response of the membrane potential at different sites of the axon when the neuron is subjected to a medium intensity electric stimulus.]{Membrane potential's response to an electric stimulus $I=100$ $\mu$A/cm$^2$ that is turned on at $t=5$ ms and turned off at $t=45$ ms, for three distinct intracellular resistances. The different colours correspond to different nodes of the axon: $x=0$ in blue, $x=1$ in orange, and $x=50$ in green.}
  \label{fig:AP-100}
\end{figure}

In the case of higher-intensity stimuli - Figure \ref{fig:AP-100} -, particularly within the unstable range of the reduced model's bifurcation diagram, the first node presents a spiking behaviour. However, the following nodes can only maintain that response for small intracellular resistances, although always slightly delayed compared with the first node. 

For R = 0.1 $k\Omega$/cm$^2$ and R = 0.5 $k\Omega$/cm$^2$, we noticed that the height and frequency of the spikes remain relatively unaffected in subsequent nodes. The primary distinction between each node's response lies in the membrane potential between each action potential, which closely approaches the resting membrane potential as we advance along the length of the axon.

However, when the intracellular resistance is sufficiently high, instead of showing a spiking pattern like the first node, the second node presents an initial action potential followed by a train of small-intensity spikes that are not transmitted to subsequent nodes. Instead, those subsequent nodes preserve only the first action potential before returning to the initial resting state. It is also relevant to note that the bigger the intracellular resistance, the later the action potential occurs in subsequent nodes.

\begin{figure}[!htb]
  \centering
  \includegraphics[width=0.99\textwidth]{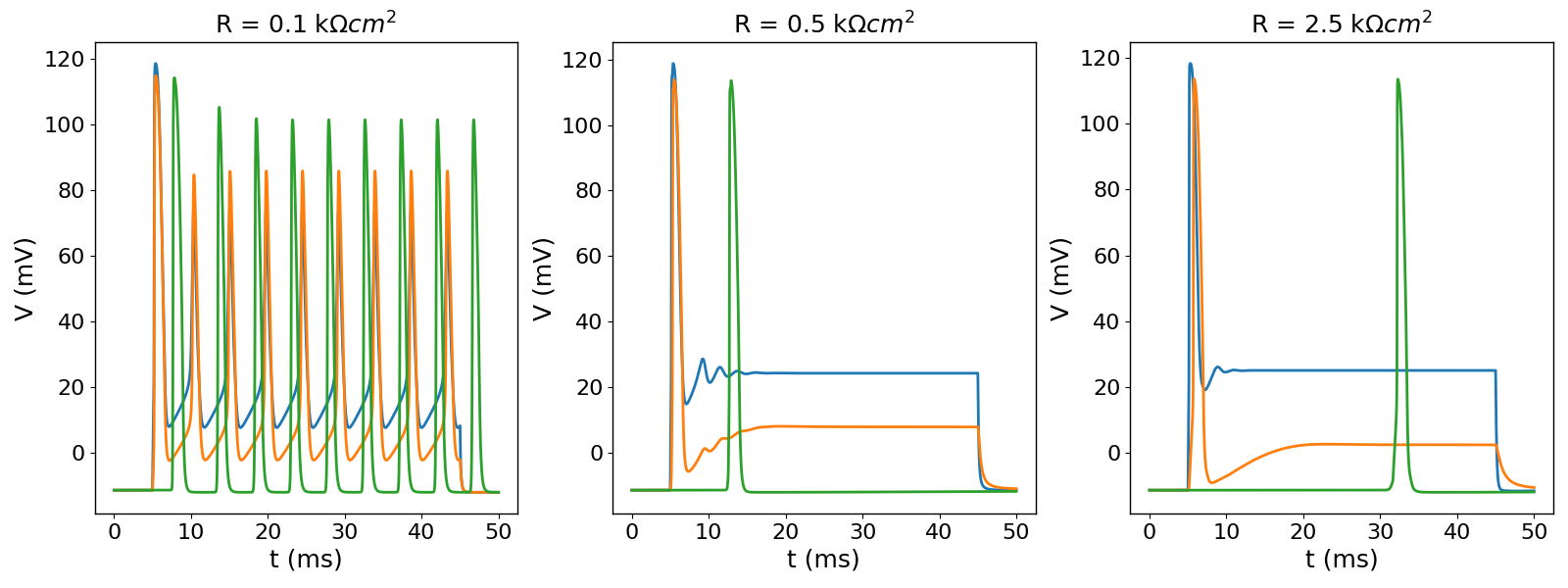}
  \caption[Response of the membrane potential at different sites of the axon when the neuron is subjected to a high intensity electric stimulus.]{Membrane potential's response to an electric stimulus $I=250$ $\mu$A/cm$^2$ that is turned on at $t=5$ ms and turned off at $t=45$ ms, for three distinct intracellular resistances. The different colours correspond to different nodes of the axon: $x=0$ in blue, $x=1$ in orange, and $x=50$ in green.}
  \label{fig:AP-250}
\end{figure}

When $I=250$ $\mu$A/cm$^2$ - Figure \ref{fig:AP-250} - it is possible, for small resistances, that the membrane on all of the $N=200$ nodes finds itself in the unstable regime. However, for the two other intracellular resistances put into test, the first node is already in the second stable regime, characterised by an action potential followed by a fast oscillatory transition to the equilibrium state. Once again, we notice that the action potential of the 51$^\text{th}$ node ($x=50$) occurs much later for R = 2.5 $k\Omega$/cm$^2$ than for R = 0.5 $k\Omega$/cm$^2$.

\subsubsection{Signal Transmission along the Axon}

The transmission of signals along the axon occurs as a peak or set of peaks in $V$ that travel through the axon. 
  
\begin{figure}[!htb]
  \begin{subfigmatrix}{3}
    \subfigure[Signal propagation for $R=0.1$ k$\Omega$cm$^2$ and I=100 $\mu$A/cm$^2$.]{\includegraphics[width=0.31\linewidth]{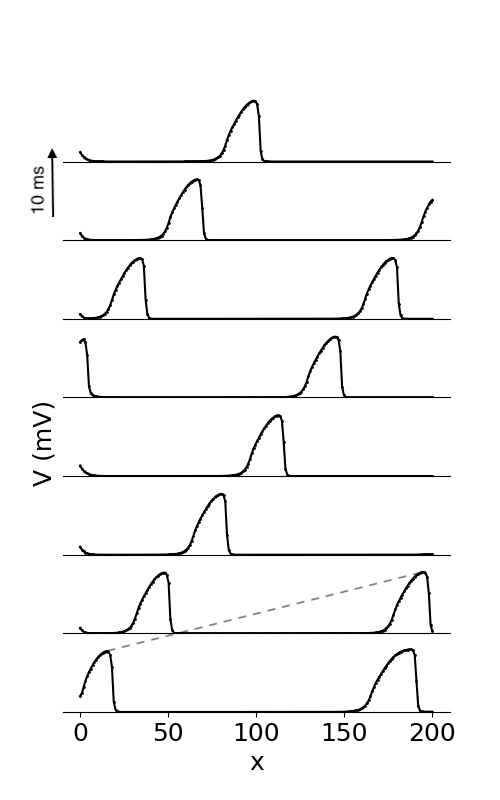}}
    \subfigure[Signal propagation for $R=0.5$ k$\Omega$cm$^2$ and I=100 $\mu$A/cm$^2$.]{\includegraphics[width=0.31\linewidth]{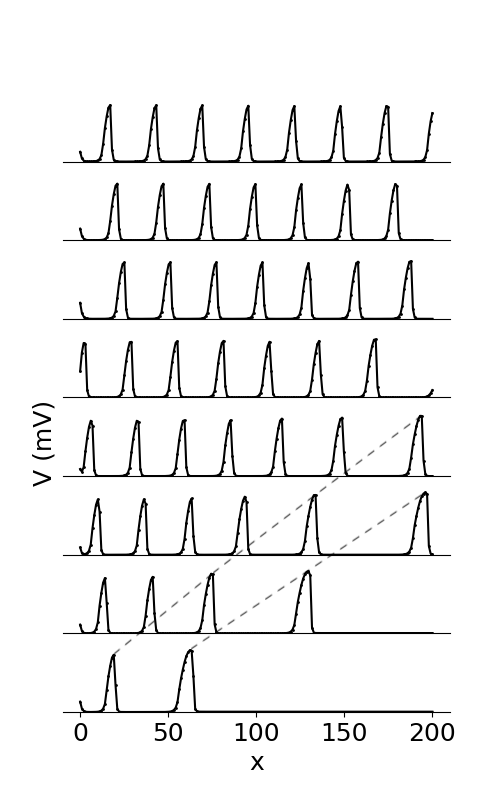}}
      \subfigure[Signal propagation for $R=5$ k$\Omega$cm$^2$ and I=100 $\mu$A/cm$^2$.]{\includegraphics[width=0.31\linewidth]{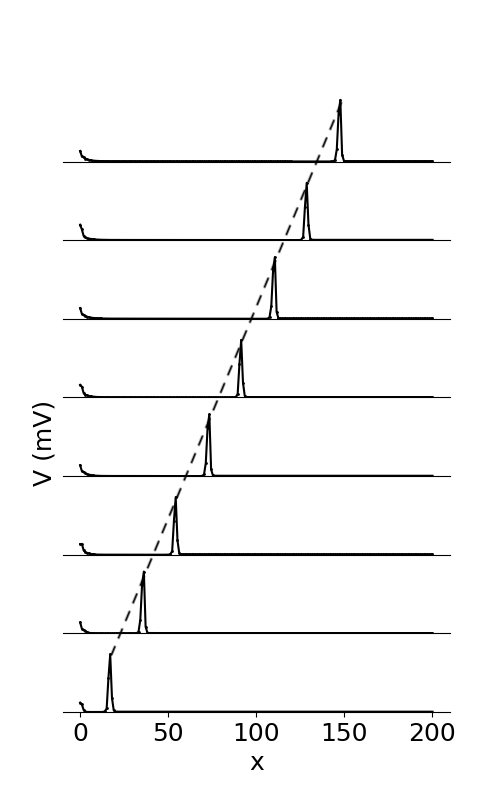}}
  \end{subfigmatrix}
  \caption[Influence of the intracellular resistance in the signal propagation along the axon.]{Influence of the intracellular resistance $R$ in the signal propagation along the axon when subjected to a stimulus I=100 $\mu$A/cm$^2$ and while considering an electrophysiological model governed by the system of equations \ref{eq:propagation}. The representation starts at the bottom of each subfigure, 10 ms after turning the electric stimulus on. The subsequent instants, 10 ms apart, are represented upwards. The vertical axis scale was fixed in all the plots, and the peak's height is approximately 100 mV in the three cases.}
  \label{fig:spatial-propagation}
\end{figure}

Figure \ref{fig:spatial-propagation} depicts the transmission of signals along the axon when $x=0$ is subjected to an external stimulus $I=100$ $\mu$A/cm$^2$. This figure comprises three subfigures, each corresponding to a particular intracellular resistance $R$. To follow the propagation of the signals over time, each subfigure includes the representation of $V(x)$ at 8 different moments,  $10$ ms apart. Starting at the bottom, we present $V(x)$ 10 ms after the stimulus $I=100$ $\mu$A/cm$^2$ is applied at $x=0$. In the top plot, we present $V(x)$ 90 ms after that instant. Although we did not aim to track the peak height on this representation, it is, as a reference, consistently close to 100 mV.

\begin{figure}[!htb]
  \centering
  \includegraphics[width=0.99\textwidth]{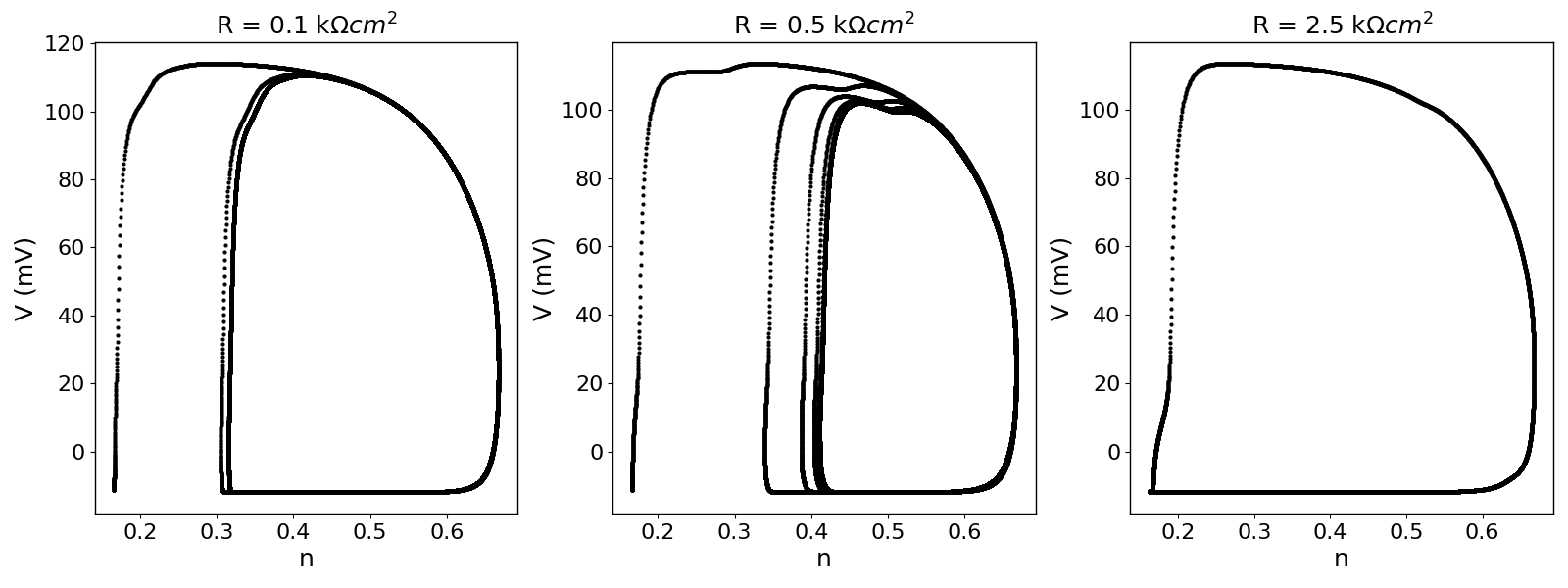}
  \caption[Influence of the intracellular resistance portrayed in the electrophysiological models' phase space for three distinct intracellular resistances.]{Influence of the intracellular resistance $R$ portrayed in the models' phase space. We considered the model ruled by the system of equations \ref{eq:propagation} and put into test three distinct intracellular resistances. The phase space observation corresponds to the evolution of $(n,V)$ at $x=50$ during 200 ms, when an electric stimulus $I=100$ $\mu$A/cm$^2$ is applied at the first node of the axon.}
  \label{fig:n(v)}
\end{figure}

Taking into account this Figure and the results obtained for a set of simulations where both $R$ and $I$ were varied in the range [0.1, 10] $k\Omega$/cm$^2$ and [1, 300] $\mu$A/cm$^2$ respectively, we concluded that the velocity $v$ of the first - and sometimes only - peak decreases with increasing resistance. However, for a fixed $R$, we observed that the velocity of the first peak displays minimal sensitivity to the intensity of the electric stimulus $I$, especially when it was the only peak propagating along the axon. In this case, the velocity remains constant irrespective of $I$. No signals were generated for $R>8$ $k\Omega$/cm$^2$, regardless of the stimulus intensity.

We have also concluded that the occurrence of one or multiple peaks only depends on the stimulus intensity that arrives at each node of the cell. If it is sufficiently high to trigger a series of spikes in all the nodes, then the signal will be transmitted as exactly that: several peaks. However, if it is too small or too high -- and we're out of the spiking regime in all the nodes --, only a peak will be transmitted throughout the axon.

When multiple peaks are transmitted along the axon -- for example as in Figures \ref{fig:spatial-propagation} a) and b) -- we found that increasing the intracellular resistance led to a decrease in the wavelength of the peaks. However, that wavelength is, initially, not regular, converging to a characteristic value for each $R$ at later times. Motivated by that finding, we examined the relationship between the membrane potential $V$ and the gating variable $n$ during the first 200 ms after applying an external stimulus $I = 100$ $\mu$A/cm$^2$ at $x=0$. In scenarios where multiple peaks were transmitted, we observed a transient behaviour in the phase space that lasts longer for higher resistance values. This effect is visible in Figures \ref{fig:n(v)} a) and b).

\begin{figure}[!htb]
  \centering
  \includegraphics[width=0.6\textwidth]{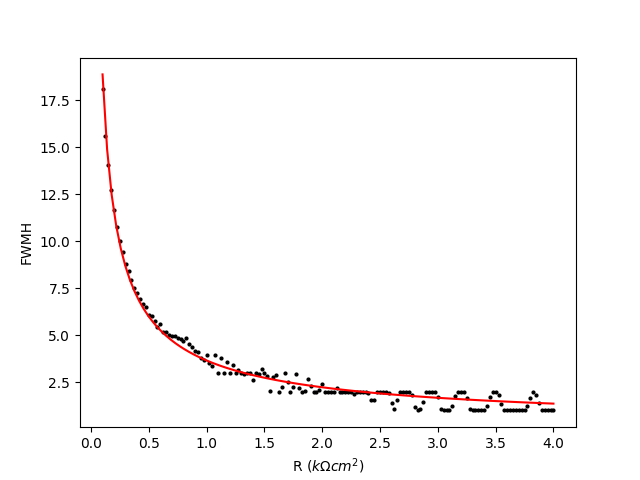}
  \caption[Relation between the full width at medium height of a signal that propagates along the axon and the neuronal intracellular resistance.]{Relation between the full width at medium height ($fwmh$) of the first peak and the intracellular resistance $R$. The data points in red correspond to the average $fwmh$ calculated for each $R$ when considering 20 external stimuli $I$ in the range [100, 200] $\mu$A/cm$^2$; the fit function of the form $fwmh = dx\:a/R^b$, with $a \approx 3.6560$ and $b\approx 0.7129$, is depicted in black. We considered the model governed by the system of equations \ref{eq:propagation}.}
  \label{fig:fwmh}
\end{figure}

Another visible finding in Figure \ref{fig:spatial-propagation} is the dependence between the width of the peak and $R$, which is, however, only minimally affected by the external stimulus $I$. Focusing on that finding, we determined the average full width at medium height ($fwmh$) for several intracellular resistances. That average was determined considering 20 external stimuli $I$ in the range [100, 200] $\mu$A/cm$^2$. The result is presented in Figure \ref{fig:fwmh}. Because of the seemingly regular dependence between the peaks' width and $R$, we fitted a function of the form $fwmh = dx\:a/R^b$ to the data and obtained $a \approx 3.6560$ and $b\approx 0.7129$.

\subsubsection{Influence of the intracellular resistance in the signal's velocity}
\begin{figure}[!htb]
  \centering
  \includegraphics[width=0.6\textwidth]{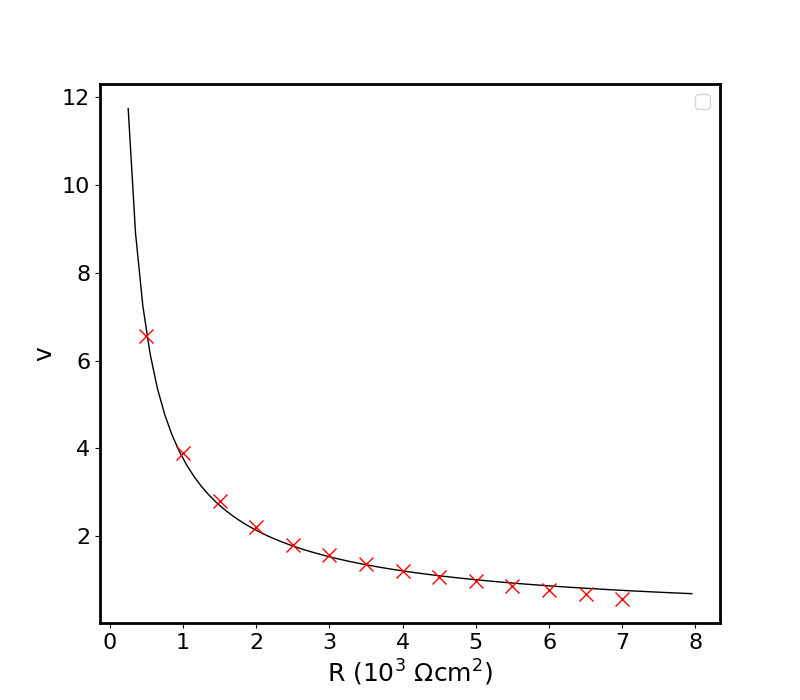}
  \caption[Relation between the propagation velocity of the signal that travels along the axon and the neuronal intracellular resistance.]{Relation between the propagation velocity $v$ and the intracellular resistance $R$. The data points in red correspond to the propagation velocity obtained when considering the model governed by the system of equations \ref{eq:propagation} with different values of $R$; in black, the fit function of the form $v =dx\: a/R^b$, with $a \approx 3.7620$ and $b \approx 0.8212$.}
  \label{fig:v(r)}
\end{figure}

To describe in more detail the dependence of the first peak's velocity on the intracellular resistance $R$, we calculated $v = \Delta x/\Delta t$, with $\Delta x$ corresponding to the number of nodes that a particular peak travels after $\Delta t = 10$ ms. To be more accurate and account for slight changes in $v$ for different external stimuli, we determined an average velocity for different intracellular resistances considering 7 different stimuli $I$: 5, 10, 20, 50, 100, 200 and 300 $\mu$A/cm$^2$. The result is depicted in Figure \ref{fig:v(r)}. 

Also in Figure \ref{fig:v(r)}, we compared the outcomes of our model with the predictions of the cable equation approach \cite{keener} by fitting a function of the form $v = dx\:a/R^b$ to the data. The cable equation predicts $b = 1/2$, with the parameter $a$ dependent on factors such as the cell capacitance $C_m$. However, our fitting results yielded $a \approx 3.7620$ and $b \approx 0.8212$, suggesting that our compartmental model is inconsistent with the cable theory assumptions. To further validate the applicability of our model and assess its performance relative to the cable equation approach, experiments on $v(R)$ can be conducted and compared to the expectations derived from our fitted function.

\subsubsection{Influence of the cell capacitance in the signal's velocity}

To examine the influence of the cell capacitance on propagation velocity, we conducted additional simulations with various values of $C_m$ ranging from 1 to 10 $\mu$F/cm$^2$, while keeping some of the previously used resistance values.

\begin{figure}[!htb]
  \centering
  \includegraphics[width=0.6\textwidth]{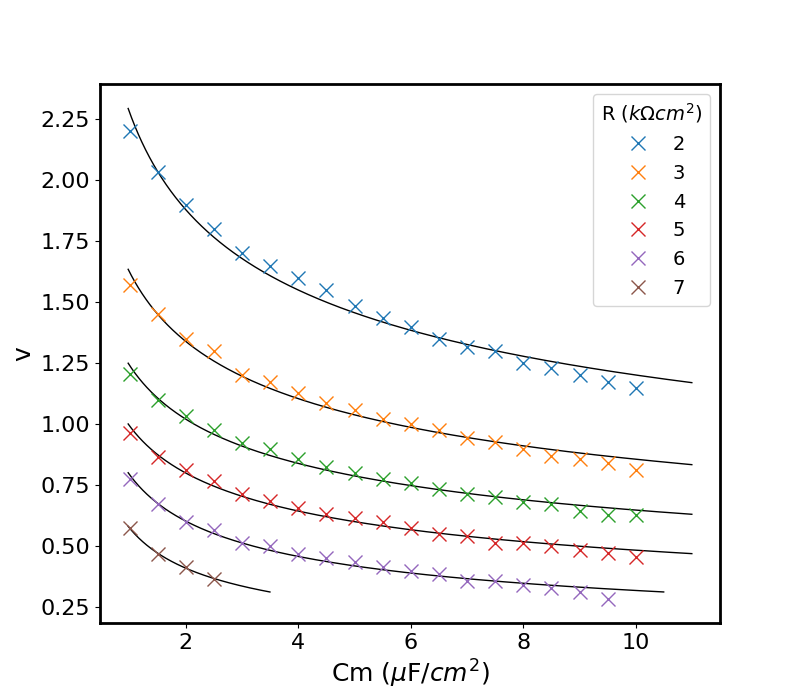}
  \caption[Relation between the propagation velocity of the electric signal that travels along the axon and the neuronal capacitance.]{Relation between the propagation velocity $v$ and the membrane's capacitance $C_m$. Once again, we considered the model described by the system of equations \ref{eq:propagation}. The data points correspond to the values of $v$ obtained for each value of capacitance and resistance put into test; in black, the fit functions of the form $v =dx\: c/C_m^d$, where the results for $c$ and $d$ are depicted in the Table \ref{tab:v(cm)}.}
  \label{fig:v(cm)}
\end{figure}

Once again, we employed a fit function of the form $v = dx\:c/C_m^d$. Based on the cable equation, we anticipate obtaining $d = 1$. However, our findings revealed that $d$ ranged from 0.28 to 0.49 for the resistances under study, and $c$ ranged from 0.57 to 2.3. The specific values of $c$ and $d$ obtained for different intracellular resistances are listed in Table \ref{tab:v(cm)} and the resulting fit functions, as well as the fitted points, are depicted in Figure \ref{fig:v(cm)}.

\begin{table}[!htb]
  \renewcommand{\arraystretch}{1.4} 
  \centering
  \begin{tabular}{lcccccc}
    \toprule
    R ($k\Omega$cm$^2$) & 2 & 3 & 4 & 5 & 6 & 7 \\
    \midrule
    c  & 2.2777 & 1.623 & 1.2411 & 0.9931 & 0.7931 & 0.5741  \\
    d  & 0.2778 & 0.2777 & 0.2822  & 0.3124 & 0.3961 & 0.4862 \\
    \bottomrule
  \end{tabular}
    \caption[Fitting results describing the relation between the propagation velocity of an electric signal and the cell capacitance for different intracellular resistances.]{Results of fitting the data points $v(C_m)$, for different intracellular resistances, to a function of the form $v = c/C_m^d$. The values of $R$ in the legend have units of $k\Omega cm^2$.}
  \label{tab:v(cm)}
\end{table}

These substantial disparities found between the compartmental and the cable models regarding the dependence of $v$ on $R$ and $C_m$ can be a starting point for experimentally evaluating which, if any, of these models is more reliable for describing neuronal signals' propagation.

\subsubsection{Dispersion Relation of neuronal signals}
We investigated the dispersion relation to further characterise our proposed signal propagation model along the axon. We focused on a cell with $R=2$ $k\Omega cm^2$ and $N=200$ to determine the wavelength and period of the first and second peaks transmitted along its length. To accomplish this, we recorded the instants at which the three first peaks crossed $x=20$ and labelled them as $t_1$, $t_2$, and $t_3$. The period of the first peak, $T_1$, was calculated as $T_1=t_2-t_1$, while the second peak period was determined as $T_2=t_3-t_2$. At $t=t_2$, we identified the position of the first peak, $x_1>20$, and designated the position of the second peak as $x_2=20$. The wavelength of the first peak, $\lambda_1$, was computed as $\lambda_1 = x_1-x_2$. Similarly, at $t=t_3$, we determined the wavelength of the second peak by calculating $\lambda_2=x_2-x_3$. It's important to note that the wavelength used here is dimensionless, representing the number of nodes between consecutive peaks at a particular instant.

\begin{figure}[!htb]
  \begin{subfigmatrix}{2}
    \subfigure[Relation between the wavelength $\lambda$ and the period $T$ for the first and second peaks.]{\includegraphics[width=0.49\linewidth]{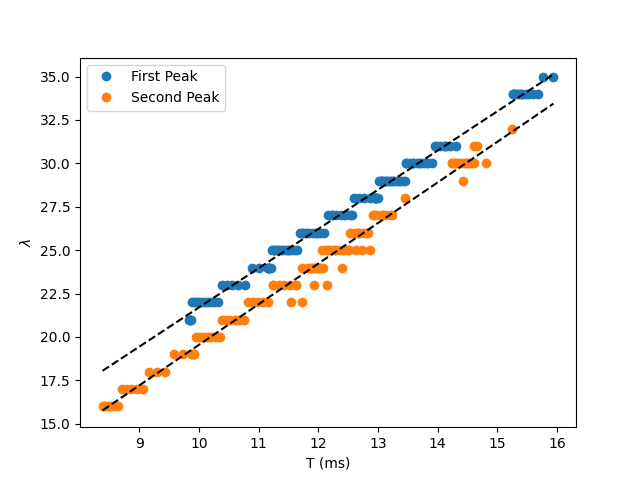}}
    \subfigure[Relation between the wave number $k$ and the radial frequency $\omega$ for the first and second peaks.]{\includegraphics[width=0.49\linewidth]{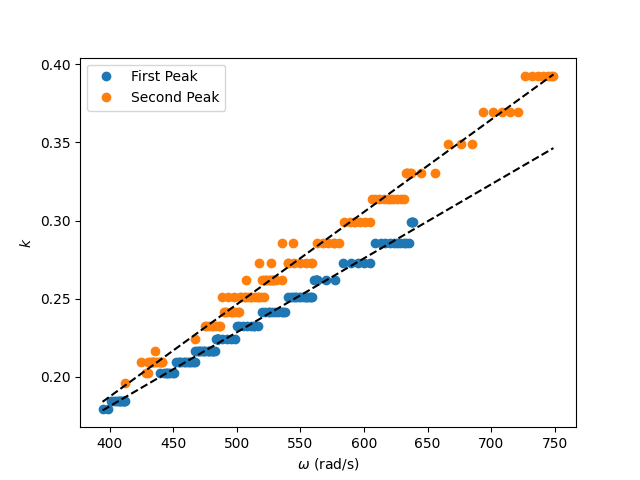}}
  \end{subfigmatrix}
  \caption[Dispersion relation of a signal that travels along an axon.]{Dispersion relation of the signals that travel along an axon's membrane with $N=100$ and intracellular resistance $R=2$ $k\Omega cm^2$, using electric stimuli in the range $[6,136]$ $\mu$A/cm$^2$. The black dashed lines result from a fit of a first-order linear function to the data.}
  \label{fig:dispersion}
\end{figure}

The obtained results for $\lambda(T)$ and $k(\omega)$ are depicted in Figure \ref{fig:dispersion}, where $k=2\pi/\lambda$ and $\omega=2\pi/T$. We considered unitary intensity stimuli in the range of $[6,136]$ $\mu$A/cm$^2$, which corresponds to a region where the signal propagates along the axon in the form of multiple peaks for this specific intracellular resistance ($R=2$ $k\Omega cm^2$)

\begin{figure}[!htb]
  \centering
  \includegraphics[width=0.7\textwidth]{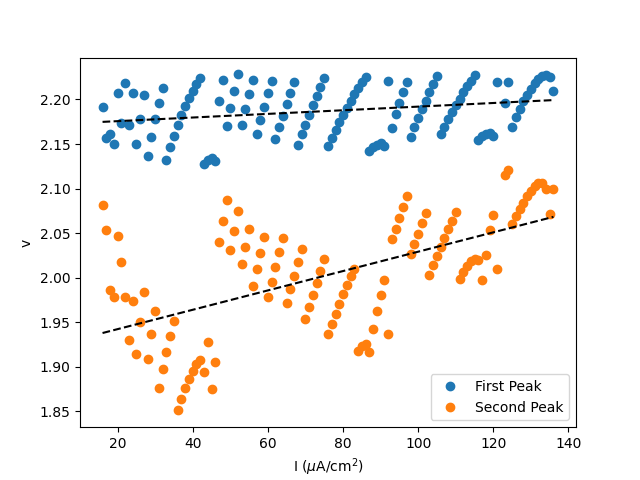}
  \caption[Velocity of the propagated signals as a function of the electric stimulus' intensity.]{Velocity $v=\lambda/T$ of the first two peaks that are propagated along an axon with $R=2$ $k\Omega cm^2$ as a function of the electric stimulus $I$ injected at $x=0$. Out of the plotted range of $I$ only one or no peaks are propagating along the axon. The black dashed lines result from a fit of a first-order linear function to the data.}
  \label{fig:v(I)}
\end{figure}

Looking at the results obtained, it is reasonable to approximate $v$ by $v=\lambda/T$. To visualise the dependence on $v=\lambda/T$ in the electric stimulus $I$ we depict, in the Figure \ref{fig:v(I)}, $v(I)$.

We start by noticing a seemingly oscillatory dependence of the peaks' velocity on the stimulus $I$. Indeed, both peaks present a characteristic pattern in $v(I)$, with moments of consistent increase in their velocity followed by an abrupt decrease at certain stimuli. However, the stimuli period for which this sudden decrease occurs is not regular. 

Both peak profiles of $v(I)$ demonstrate a bottleneck effect, with a narrowing range of velocity values as stimulus intensity increases. The first peak maintains a relatively small oscillation amplitude around its average velocity, while the second peak exhibits a noticeable velocity increase with increasing stimulus intensity. This distinction is particularly evident when fitting the data with first-order linear functions, represented by the black dashed lines in Figure \ref{fig:v(I)}.

\subsubsection{Membrane's Response to Random Perturbations from Resting State}

Although spatial uniformity can be achieved experimentally in vivo, the intricate structure of the neurone and the non-uniform distribution of charged bodies in the intracellular space create spatial gradients in the membrane potential under natural conditions.

Therefore, our next step involved evaluating the models' response to potential perturbations from the equilibrium state $V_0 = -11.3554$ mV. To accomplish this, we explored a set of different scenarios, which are illustrated in the upcoming figures.

\begin{figure}[!htb]
  \begin{subfigmatrix}{2}
    \subfigure[Random perturbation at two nodes]{\includegraphics[width=0.49\linewidth]{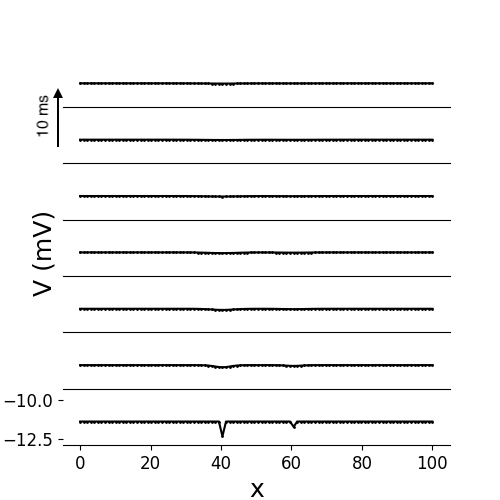}}
    \subfigure[Random perturbation at all nodes]{\includegraphics[width=0.49\linewidth]{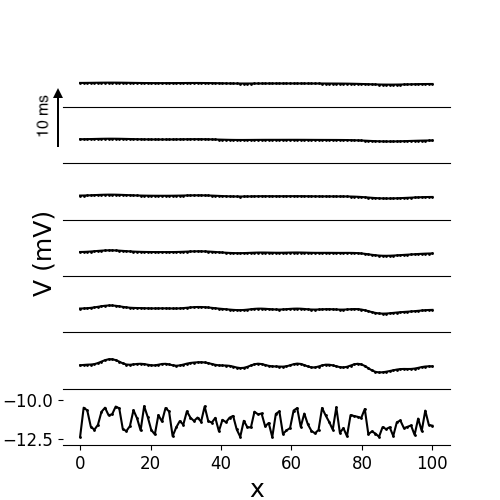}}
  \end{subfigmatrix}
  \caption[Evolution of random perturbations on the axon's membrane potential.]{Evolution of random perturbation on the resting potential of an axon membrane with $N=100$ and intracellular resistance $R=4$ $k\Omega cm^2$. The representation starts at the bottom of each subfigure ($t=0$), where the initial perturbation can be observed. The consequent response/evolution of the membrane's potential along the length of the axon can be observed throughout the following 80 ms in the upper plots, which are 10 ms apart from each other.}
  \label{fig:simple-perturbation}
\end{figure}

In Figure \ref{fig:simple-perturbation}, we depict a scenario where no stimulus was injected at $x=0$, enabling us to observe whether the membrane would smoothly return to its equilibrium state or propagate - or even amplify - the perturbations. The inflicted perturbations were random, causing the membrane's potential to deviate from $V_0$ to some value within the range of $V_0-1$ mV to $V_0+1$ mV. We examined two cases: one where perturbations were applied at $x=40$ and $x=60$ (depicted in Figure \ref{fig:simple-perturbation} (a)), and other where all nodes of the membrane were randomly perturbed (depicted in Figure \ref{fig:simple-perturbation} (b)). The bottom plot of each subfigure displays the membrane's potential profile at $t=0$, while the upper plots illustrate its evolution in intervals of 10 ms since $t=0$. The vertical axis scale - depicting the region around $V_0 = -11.3554$ mV - is the same in all the plots.

In both cases, we concluded that the perturbations - localised or not - disappear rapidly and without causing any impact on the cell membrane. In fact, and as we confirmed for other values of intracellular resistances, the membrane always restores its equilibrium state $V_0$ along its length, eliminating the perturbation.

\begin{figure}[!htb]
  \begin{subfigmatrix}{2}
    \subfigure[Long lasting electric stimulus]{\includegraphics[width=0.49\linewidth]{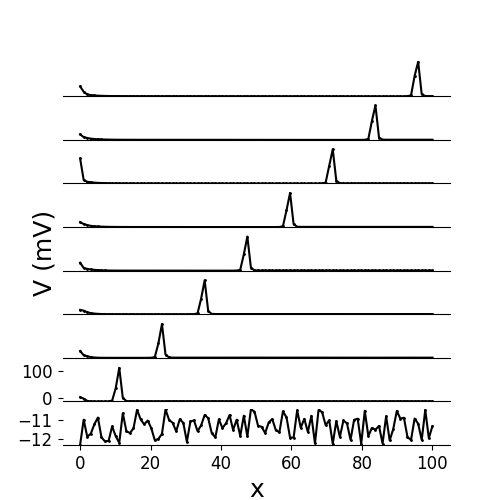}}
    \subfigure[Short pulse electric stimulus]{\includegraphics[width=0.49\linewidth]{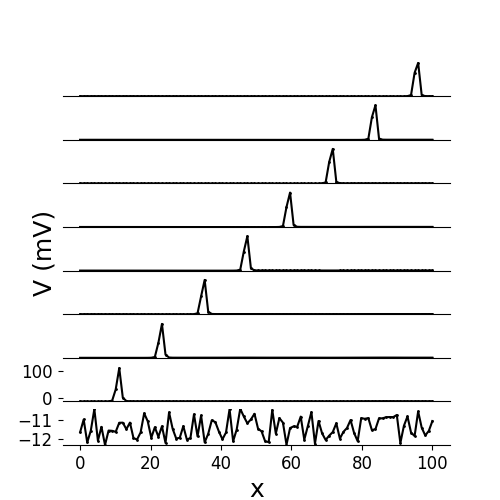}}
  \end{subfigmatrix}
  \caption[Signal propagation along an axon with perturbed initial membrane potential.]{Random perturbation of the resting potential of an axon with and $R=4$ $k\Omega cm^2$ and consequent response throughout the following 80 ms when subjected to an external stimulus at $x=0$ and $t=0$. In a), we considered a stimulus $I = 100$ $\mu$A/cm$^2$ that was active (stimulating the first node of the axon) throughout the entire simulation. In b), we considered that the same intensity stimulus $I= 100$ $\mu$A/cm$^2$ was only active for 1 ms, being shut down at $t=1$ ms. }
  \label{fig:perturbation}
\end{figure}

Due to the likelihood of a real neurone experiencing slight and random deviations from its resting potential along its length, we investigated the effects of these random perturbations in the presence of external stimulus at $x=0$. A representative result of these tests is illustrated in Figure \ref{fig:perturbation}. In this case, a cell with resistance $R=4$ $k\Omega cm^2$ was subjected to an external electric stimulus $I = 100$ $\mu$A/cm$^2$ at $x=0$. In Figure \ref{fig:perturbation} a), that stimulus was only maintained for 1 ms. In Figure \ref{fig:perturbation} b), the stimulus was maintained for 100 ms. At $t=0$, the membrane potential assumed random values on the $N=100$ nodes within the range [$V_0-1$,$V_0+1$] mV. In these two figures, the scale of the bottom plot, which represents the membrane potential profile at $t=0$ and around $V_0 = -11.3554$ mV, differs from the upper plots, which maintain the same scale as depicted at $t=10$ ms to enhance the visualisation of the peak's propagation. We plotted the axon's potential profile for the first 80 ms, which approximately represents the time it takes for the signal to propagate from one end to the other of the axon with $N=100$ nodes and $R=4$ $k\Omega cm^2$.

Once again, our findings indicate that these perturbations of the cell membrane have no significant effects on the propagation of signals. For the multiple values of intracellular resistances and electric stimuli put into test (resistances in the range $R\in[0.1, 7]$ $k\Omega cm^2$ and $I\in[5, 250]$ $\mu$A/cm$^2$, the propagation velocity remained unchanged compared to previous tests conducted without any perturbations. We also concluded that, despite lasting only 1 ms, the external stimulus $I = 100$ $\mu$A/cm$^2$ could still propagate through the axon, as we would expect from a real neurone.

\subsubsection{Transitioning from Discretised to Continuous Compartmental Model}
To determine the continuous version of this compartmental model, we start by considering that the distance between each node, $dx$ becomes infinitely small compared to the dimension of the axon. Moreover, we also assume that $V$ is a sufficiently smooth function of $x$ and, thus, that we can write the membrane potential at neighbouring points of the node $j$ as their respective second-order Taylor expansion:

\begin{equation*}
\begin{aligned}
	V_{j-1} = V_j -  dx \frac{\partial V}{\partial x} + \frac{dx^2}{2} \frac{\partial^2 V}{\partial x^2}\\
	V_{j+1} = V_j + dx \frac{\partial V}{\partial x} + \frac{dx^2}{2} \frac{\partial^2 V}{\partial x^2}
	\end{aligned}
\end{equation*}

Replacing $V_{j-1}$ and $V_{j+1}$ in the equation \ref{prop-model}, by their expansions around $V_j$, we obtain

\begin{equation}
	\begin{aligned}
	C_m \frac{\partial V}{\partial t} = I + \frac{dx}{R}\frac{\partial V}{\partial x} - I_{ion}(V, n), \:\:  x = 0 \\
	C_m \frac{V}{dt} = \frac{dx^2}{R}\frac{\partial^2 V}{\partial x^2}- I_{ion}(V, n), \:\: 0 < x < L \\
	C_m \frac{\partial V}{\partial t} = - \frac{dx}{R}\frac{\partial V}{\partial x} - I_{ion}(V, n), \:\:  x = L \\
	\frac{dn}{dt} =\alpha_n(V)(1-n)-\beta_n(V)n
    \end{aligned}
    \label{eq:prop-final-model}
\end{equation}

considering that $\frac{\partial V}{\partial x}$ is much greater than $\frac{dx^2}{2}\frac{\partial^2 V}{\partial x^2}$ and that the length of the axon is $L$. Equation \ref{eq:prop-final-model}b) is a nonlinear reaction-diffusion equation with new types of boundary conditions. The boundary condition at $x=0$ characterises the transmission of an electric stimulus from the soma to one side of the axon, while the second boundary condition represents the arrival of the transmitted neuronal signal at the synaptic termination of the axon. Together, this set of equations describes the propagation of non-uniform diffusion waves \cite{canodilao2} along the axon.
\cleardoublepage

\cleardoublepage

\chapter{Conclusions}
\label{chapter:conclusions}

In this work, we comprehensively analysed the Hodgkin-Huxley model to understand better the precise roles of each ionic channel in the membrane's response to external stimuli. Our findings indicate that the Potassium channel predominantly contributes to the stability of the cell membrane, while the inclusion of Sodium channels leads to unstable behaviour when subjected to intermediate-intensity stimuli. On the other hand, the impact of the Leak channel in the features of the system is minimal, resulting in a slight shift in the equilibrium state of the membrane when subjected to low-intensity stimuli. As a result, we excluded the leak channel on subsequent tests and simulations, eliminating two unnecessary parameters.

Regarding the gating variables of the HH model, we found that $h$ of the Sodium channel reflects the response of $n$. We extensively tested different values and concluded that setting $c$ to 0.71 yielded the closest results to the HH model's bifurcation diagram without the leak channel when replacing $h$ by $c-n$. Additionally, due to its rapid response to external stimuli, we found that approximating $m$ by its asymptotic value in the reduced model did not compromise the essential features of the original Hodgkin-Huxley approach. Concluding this model reduction, we obtained an electrophysiological model with only two independent variables, which provided a much clearer analysis of the system's behaviour with improved numerical efficiency.

To better characterise how different parameters influence neuronal behaviour in the unstable regime, we analysed the spiking frequency $f_s$ of cells with different characteristics when subjected to several intensity stimuli. We concluded that, in the unstable regime, all cells present an increase in the spiking frequency as the stimulus intensity is raised. We also observed that decreasing $g_{Na}$, $V_{K}$ and $V_{Na}$ all contributed to a decrease in the spiking frequency for a given stimulus while decreasing $g_K$ had the opposite effect. From these parameters, it was also left clear that the membrane is more sensitive to changes in the features of the Potassium channel.

The capacitance of the membrane also proved to have a non-negligible effect on the spiking frequency, as reducing $C_m$ resulted in a considerable increase in $f_s$. Moreover, cells with a limited capacity to store charge exhibited modifications in the bifurcation diagram of the system, particularly near the second Hopf bifurcation point. For $C_m = 0.01$ $\mu$F/cm$^2$, in particular, an unstable limit cycle emerged from that point, and we observed the presence of two types of Canards near $I = 217.697$ $\mu$A/cm$^2$.

Another objective of our investigation was to develop a model for signal propagation along the axon. We proposed a compartmental model in which all nodes had the same capacitance and types of ionic channels, incorporating our previously developed reduced electrophysiological model. We examined the response of the membrane potential along the axon under different conditions and analysed the transmission of signals along the axon for varying-intensity stimuli. Our findings indicated that signal propagation could occur as single or multiple peaks. In scenarios with multiple peaks, we observed a transient behaviour in the phase space $(V, n)$, which persisted longer for higher resistance values.

Furthermore, we thoroughly characterised the influence of intracellular resistance $R$ on the features of the signal, including the full width at medium height and velocity. Increasing $R$ resulted in narrower signals and decreased peak velocity, approximately following the relationships $fwmh = 3.6/R^{-0.7}$ and $v = 3.8/R^{0.8}$ for a cell with $C_m = 1$ $\mu$F/cm$^2$. Considering that we had observed the significant impact of $C_m$ on the membranar response to external stimuli, we also examined the relationship between peak velocity and capacitance for cells with different intracellular resistances. Our analysis demonstrated that signals propagate faster in cells with a smaller capacity to store electric charge.

Continuing our investigation, we studied the dispersion relation of neuronal signals in scenarios where signal propagation occurred in the form of multiple peaks and we found an approximately linear relation between the wavelength and the frequency of the first and second peaks. After calculating the velocities of the first two peaks as $v=\lambda/T$, we observed an oscillatory dependence of $v$ on the stimulus intensity $I$. However, that dependence was found to be substantially different between the first and second peaks, with the first maintaining a relatively small amplitude of oscillation around its average velocity and the second exhibiting a noticeable overall velocity increase with increasing stimulus intensity.

We then proved that this spatial model is not affected by perturbations in the membrane potential along the axon and presented a reaction-diffusion model that describes the propagation of non-uniform unidirectional diffusion waves along the axon.

In the future, it would be interesting to explore the potential applicability of this compartmental model in describing signal propagation in the heart. Such investigations could catalyse studying cardiovascular pathologies and arrhythmias, enhancing our understanding of these conditions.

\cleardoublepage

\phantomsection
\addcontentsline{toc}{chapter}{\bibname}

\bibliographystyle{plain}
\bibliography{Thesis_bib_DB}

\cleardoublepage

\end{document}